\documentclass[11pt,english,twoside]{article}

\usepackage[T1]{fontenc}
\usepackage[latin1]{inputenc}
\usepackage[english]{babel}
\usepackage{lmodern}
\usepackage{a4wide}
\usepackage{amssymb, amsmath, amsthm}
\usepackage{slashed}
\usepackage{float}
\usepackage{graphicx}
\usepackage[dvips]{epsfig}
\usepackage{psfrag}
\usepackage{lscape}
\usepackage[all]{xy}
\usepackage{hyperref}
\usepackage{enumerate}
\usepackage{dsfont}
\usepackage{color}
\usepackage{mathabx}
\usepackage{mathtools}
\usepackage{framed}

\newcommand{\beq}{\begin{equation}}
\newcommand{\eeq}{\end{equation}}
\def\bea#1\eea{\begin{align}#1\end{align}}
\newcommand{\nn}{\nonumber}
\newcommand{\w}{\wedge}
\newcommand{\id}{\mathds{1}}
\newcommand{\ov}{\overline}
\renewcommand{\i}{\ensuremath{\textnormal{i}}}

\def\del {\partial}
\def\d {{\rm d}}
\def\R {\mathcal{R}}

\def\hhh {\mathcal{H}}

\def\tg {\tilde{g}}
\def\b {\beta}
\def\tp {\tilde{\phi}}

\def\te {\tilde{e}}

\def\p {\phi}
\def\P {\Phi}
\def\cR {\widecheck{\cal R}}

\def\cN {\widecheck{\nabla}}
\def\g {\gamma}
\def\G {\Gamma}
\def\o {\omega}
\def\co {\widecheck{\omega}}

\def\eps {\epsilon}
\def\la {\lambda}

\def\N {\nabla}
\def\al {\alpha}
\def\T {\mathcal {T}}

\def\D {\mathcal{D}}
\DeclareMathOperator{\re}{Re}
\DeclareMathOperator{\im}{Im}

\def\mmm {{\cal M}}
\def\eee {\mathcal{E}}
\def\teee {\tilde{\mathcal{E}}}

\newcommand{\s}{\slashed}

\newcommand{\sla}{\slash\!\!\!\!}

\makeatletter
\@addtoreset{equation}{section}
\makeatother

\begin{document}

\begin{titlepage}

\begin{center}

\rightline{\small AEI-2014-060}

\rightline{\small MPP-2014-555}

\vskip 2cm

{\fontsize{16.1}{21}\selectfont \noindent\textbf{Supersymmetry with non-geometric fluxes, or a $\b$-twist \\ \vskip 0.25cm in Generalized Geometry and Dirac operator} }

\vskip 2.1cm

\textbf{David Andriot${}^{a,b}$, Andr\'e Betz${}^{c}$}

\vskip 0.6cm

\begin{enumerate}[$^a$]
\item \textit{Max-Planck-Institut f\"ur Gravitationsphysik, Albert-Einstein-Institut,\\Am M\"uhlenberg 1, 14467 Potsdam-Golm, Germany}
\vskip 0.1cm
\item \textit{Institut f\"ur Mathematik, Humboldt-Universit\"at zu Berlin, IRIS-Adlershof,\\Zum Gro\ss en Windkanal 6, 12489 Berlin, Germany}
\vskip 0.1cm
\item \textit{Max-Planck-Institut f\"ur Physik,\\F\"ohringer Ring 6, 80805 M\"unchen, Germany}
\end{enumerate}

\vskip 0.2cm

{\small \texttt{david.andriot@aei.mpg.de}, \texttt{abetz@mpp.mpg.de}}

\end{center}

\vskip 2.0cm

\begin{center}
{\bf Abstract}
\end{center}

\noindent We study ten-dimensional supersymmetric vacua with NSNS non-geometric fluxes, in the framework of $\b$-supergravity. We first provide expressions for the fermionic supersymmetry variations. Specifying a compactification ansatz to four dimensions, we deduce internal Killing spinor equations. These supersymmetry conditions are then reformulated in terms of pure spinors, similarly to standard supergravity vacua admitting an SU(3)$\times$SU(3) structure in Generalized Complex Geometry. The standard $\d-H\wedge$ acting on the pure spinors is traded for a generalized Dirac operator $\D$, depending here on the non-geometric fluxes. Rewriting it with an exponential of the bivector $\b$ leads us to discuss the geometrical characterisation of the vacua in terms of a $\b$-twist, in analogy to the standard twist by the $b$-field. Thanks to $\D$, we also propose a general expression for the superpotential to be obtained from standard supergravities or $\b$-supergravity, and verify its agreement with formulas of the literature. We finally comment on the Ramond-Ramond sector, and discuss a possible relation to intermediate or dynamical SU(2) structure solutions.

\vfill

\end{titlepage}

\tableofcontents

\newpage

\section{Introduction and main results}

An important problem in string phenomenology is the presence of massless, unobserved, scalar fields in four-dimensional effective theories derived from string theory: the moduli. Considering fluxes in the ten-dimensional background on which one compactifies usually helps, since they contribute to the four-dimensional potential and can thus stabilise moduli. A motivation to study non-geometric fluxes is the hope of using them for the same purpose: they were shown in several examples to stabilise moduli or to lead to de Sitter vacua \cite{Shelton:2006fd, Micu:2007rd, Palti:2007pm, deCarlos:2009qm, Danielsson:2012by, Blaback:2013ht, Damian:2013dq, Damian:2013dwa, Dall'Agata:2012sx, Catino:2013syn}. These fluxes initially appeared in four-dimensional gauged supergravities: they correspond to some gaugings or components of the embedding tensor \cite{Shelton:2005cf, Dabholkar:2002sy, Dabholkar:2005ve}. However, they cannot be obtained by a compactification from a standard supergravity. They are rather traditionally considered to descend from ten-dimensional non-geometric backgrounds \cite{Hellerman:2002ax, Dabholkar:2002sy, Flournoy:2004vn}. Their derivation was clarified recently, at least for a class of backgrounds, thanks to the formalism of $\b$-supergravity \cite{Andriot:2013xca, Andriot:2014uda} (reviews on these ideas can be found in \cite{Wecht:2007wu, Andriot:2011uh, Andriot:2013txa}). This ten-dimensional local reformulation of standard supergravity has two important features: it depends explicitly on non-geometric fluxes, giving them a manifest ten-dimensional origin, and a non-geometric background of standard supergravity can be reformulated into a geometric one of $\b$-supergravity, allowing then for a compactification. Some vacua of four-dimensional gauged supergravities with non-geometric fluxes thus get a clear ten-dimensional uplift. Now, one can hope to find ten-dimensional backgrounds with non-geometric fluxes that would stabilise moduli after compactification. In this paper, we are interested in supersymmetric backgrounds of $\b$-supergravity and their geometrical characterisation, as detailed below.

The NSNS sector of $\b$-supergravity is obtained by the field redefinition \eqref{fieldredef} from the standard supergravity metric $g_{MN}$, $b$-field and dilaton $\p$ to a new metric $\tg_{MN}$, an antisymmetric bivector $\b^{MN}$ and a new dilaton $\tp$. This is equivalent to parameterizing as in \eqref{genvielb} the generalized metric $\hhh$ with the generalized vielbein $\teee$ instead of $\eee$ \cite{Andriot:2011uh}. The new fields allow to define ten-dimensional NSNS non-geometric $Q$- and $R$-fluxes in flat indices
\beq
Q_{C}{}^{AB} = \del_C \b^{AB} - 2 \b^{D[A} f^{B]}{}_{CD}\ ,\quad R^{ABC} = 3 \b^{D[A}\N_D \b^{BC]} \ , \label{fluxesintro}
\eeq
as in \cite{Grana:2008yw, Aldazabal:2011nj, Andriot:2012an}. Building on \cite{Andriot:2011uh, Andriot:2012wx, Andriot:2012an}, the Lagrangian and equations of motion for the NSNS sector of $\b$-supergravity were derived in terms of these fluxes in \cite{Andriot:2013xca, Andriot:2014uda}. A completion of the theory to other sectors should be possible by further reformulating standard supergravities, as discussed in \cite{Andriot:2013xca} (see here section \ref{sec:RR} on RR fluxes). The theory is also expected to retain symmetries of standard supergravities, such as supersymmetry (SUSY). But we do not have so far the fermionic fields and their SUSY variations. Natural candidates for the latter can nevertheless be derived from the Generalized Geometry formulation of $\b$-supergravity, obtained in \cite{Andriot:2013xca}. For standard type II supergravities, it was noticed \cite{Coimbra:2011nw} that the SUSY variations of the gravitini $\psi^{1,2}_{M}$ and dilatini $\la^{1,2}$ could be rephrased in terms of Spin(9,1)$\times$Spin(1,9) derivatives, as in \eqref{fermvarSpin}. These spinorial derivatives are determined independently, applying the Generalized Geometry formalism, and we derived them for $\b$-supergravity. They allow as well to reproduce the Lagrangian and the equations of motion, as was checked both for standard and $\b$-supergravity. We thus consider that they should play an analogous role in both theories for the SUSY variations as well, and infer for fermions of $\b$-supergravity
\bea
\delta \tilde{\psi}^{1,2}_{M}=& \te^A{}_M \left( \N_{A} \pm \eta_{AD} \cN^{D} - \frac{1}{8} \eta_{AD} \eta_{BE} \eta_{CF} R^{DEF} \G^{BC} \right) \eps^{1,2} \label{fermvarintro}\\
\delta \tilde{\rho}^{1,2}=&\left( \G^A \N_A \mp \G^A \eta_{AD} \cN^D  + \frac{1}{24} \eta_{AD} \eta_{BE} \eta_{CF} R^{DEF} \G^{ABC} - \G^A \del_A \tp \mp \G^A \eta_{AB} (\b^{BC} \del_C \tp - \T^B) \right) \eps^{1,2} \ ,\nn
\eea
(only the NSNS contribution) with $\tilde{\rho}^{1,2} \equiv \G^A \tilde{\psi}^{1,2}_{A}-\tilde{\la}^{1,2}$, and $\eps^{1,2}$ the SUSY fermionic parameters. The upper/lower sign refers to the first/second number (1/2), $\cN^A$ is a covariant derivative whose spin connection depends on the $Q$-flux, and further notations are detailed in the paper. These fermionic SUSY variations will be our starting point in what follows.

When looking for new vacua of a theory, SUSY is a useful tool that provides technical simplifications. In addition, the conditions for a vacuum to preserve SUSY bring constraints that characterise, in particular geometrically, the allowed backgrounds. One gets this way information on the properties of the possible vacua. This motivates the present work for $\b$-supergravity. These features are well illustrated for standard ten-dimensional type II supergravities in \cite{Grana:2005sn, Grana:2006kf} that we follow here. For a vacuum to preserve SUSY, the fermionic SUSY variations should vanish. One further considers a compactification ansatz for the background: in particular the geometry is a warped product of a four-dimensional Minkowski (Mink) or Anti de Sitter (AdS) space-time, and an internal six-dimensional manifold $\mmm$. The conditions for SUSY are then decomposed accordingly. For (at least) ${\cal N}=1$ SUSY in four dimensions, the SUSY conditions were reformulated in \cite{Grana:2005sn} in terms of Generalized Complex Geometry (GCG) \cite{Hitchin:2004ut, Gualtieri:2003dx}, into the pure spinors conditions \eqref{dP1stand} and \eqref{dP2stand}. The mathematical framework of GCG provides interesting interpretations of the supergravity quantities and leads to a geometrical characterisation of $\mmm$ for a Mink SUSY vacuum: for $\mu=0$ ($\mu$ is related to the four-dimensional cosmological constant), the condition \eqref{dP1stand} indicates that $\mmm$ has to be a twisted generalized Calabi-Yau (GCY). This generalizes the Calabi-Yau (CY) condition, valid in absence of background fluxes. The pure spinors conditions are given by
\bea
\!\!\!\! e^{\p}\left(\d -  H\w \right) \left(e^{-\p} \P_1\right) + e^{-2A}& \d (e^{2A}) \w \P_1 =2\varepsilon\ e^{-A} \mu\ \re(\Phi_2) \label{dP1stand}\\
\!\!\!\! e^{\p}\left(\d -  H\w \right) \left(e^{-\p} \P_2\right) + e^{-2A}& \d (e^{2A}) \w \P_2 =3\varepsilon\ e^{-A}\ \i \im(\overline{\mu} \Phi_1) + e^{-A}\d (e^{A}) \w \ov{\P_2} + \mbox{RR} \ , \label{dP2stand}
\eea
with the flux $H=\d b$, the warp factor $e^{2A}$, the Ramond-Ramond contribution RR, and
\beq
\mbox{IIA} :\ \P_1=\P_+\ ,\ \P_2=\P_-\ ,\ \varepsilon=+1\ ,\qquad \mbox{IIB} :\ \P_1=\P_-\ ,\ \P_2=\P_+\ ,\ \varepsilon=-1 \ .\label{IIAB}
\eeq
$\Phi_{\pm}$ are pure spinors in GCG, they are in particular O(6,6) spinors, but can be viewed also as polyforms: for example, a (simplified) expression is given for them in \eqref{purespinSU3simpleintro}, in the case of an SU(3) structure.\footnote{The conditions for a vacuum to preserve SUSY in various theories have been reformulated into analogues of the pure spinors conditions, applying similar techniques to space(-time)s of dimension D (the external part being Minkowski or Anti de Sitter). For type II and M-theory, such conditions were obtained for D from 11 down to 3 in \cite{Prins:2014ona, Tomasiello:2011eb, Prins:2013koa, Rosa:2013lwa, Prins:2013wza, Smyth:2009fu, Haack:2009jg, Gabella:2012rc, Gabella:2009wu, Lust:2010by, Apruzzi:2014qva, Apruzzi:2013yva}. Rewritings of the previous conditions were worked-out in \cite{Tomasiello:2007zq, Rosa:2013jja, Giusto:2013rxa}. The heterotic case (${\cal N}=1$ Mink vacuum) was treated in \cite{Andriot:2009fp}. Finally conditions for type II Mink vacua were also written in terms of Exceptional Generalized Geometry (EGG) in \cite{Grana:2009im, Grana:2011nb, Grana:2012ea}.} In this paper, we follow the same procedure, described in \cite{Grana:2006kf}, and derive analogous pure spinors conditions for $\b$-supergravity (including only the NSNS sector)
\bea
\frac{1}{2} \D \P_1 + e^{-2A}& \left( \d  + \cN^a \cdot \iota_a \right)(e^{2A}) \P_1 =2\varepsilon\ e^{-A} \mu\ \re(\Phi_2) \label{dP1intro}\\
\frac{1}{2} \D \P_2 + e^{-2A}& \left( \d  + \cN^a \cdot \iota_a \right)(e^{2A}) \P_2 =3\varepsilon\ e^{-A}\ \i \im(\overline{\mu} \Phi_1) + e^{-A}\left( \d  - \cN^a \cdot \iota_a \right)(e^{A}) \ov{\P_2} \ , \label{dP2intro}
\eea
where $\D$ is an important operator given in \eqref{Diraccurved}, and the other derivatives act only on the warp factor. On the contrary to \eqref{dP1stand} and \eqref{dP2stand}, the conditions \eqref{dP1intro} and \eqref{dP2intro} are necessary but not sufficient to preserve SUSY, as shown in appendix \ref{ap:suff}. This is probably due to the absence of a RR contribution. The latter is expected to be simply added to the right-hand side (RHS) of \eqref{dP2intro}, so the present work remains relevant.

The $\D$ obtained in the two conditions \eqref{dP1intro} and \eqref{dP2intro} turns out to be precisely the generalized Dirac operator computed in \cite{Andriot:2014uda}. There, we considered $\D =\G^{{\cal A}} D_{{\cal A}}$ where $D_{{\cal A}}$ is the Spin(d,d)$\times \mathbb{R}^+$ covariant derivative in a d-dimensional space-time \cite{Grana:2008yw, Hohm:2011dv, Coimbra:2011nw, Jeon:2011vx, Jeon:2012kd, Jeon:2012hp, Geissbuhler:2013uka}, and we chose for the Cliff(d,d) $\G^{{\cal A}}$-matrices a representation as forms and contractions. The generalized spin connection components were already determined in terms of fluxes in \cite{Andriot:2013xca}, where we used the generalized torsion free condition \cite{Coimbra:2011nw} and further fixing, starting with the generalized vielbein $\teee$ \eqref{genvielb} for $\b$-supergravity. Proceeding similarly with $\eee$ for standard supergravity, we derived in \cite{Andriot:2014uda} both Dirac operators $\D$ acting on a form $A_p$
\bea
& \ \ \mbox{for standard supergravity:} \quad \D A_p = 2 e^{\p}\left(\d -  H\w \right) \left(e^{-\p} A_p \right) \label{Diracstandard}\\
& \phantom{\quad \D A_p } \qquad  \quad = 2\left( \del_a \cdot e^a\! \w - f \diamond  - H \w - \d \p \w \right) A_p \ ,\label{Diracflatstandard}\\
& \ \ \mbox{for $\b$-supergravity:} \quad \D A_p = 2 e^{\tp}\left(\d -  \cN^a \cdot \iota_a+\T \vee + R \vee\right) \left(e^{-\tp} A_p \right) \label{Diraccurved}\\
& \phantom{\quad \D A_p } \qquad  \quad = 2 \left(\del_a\cdot \te^a\!\w + \b^{ab}\del_b\cdot\iota_{a}- f \diamond - Q \diamond + R \vee  -\d \tp \w + (\cN \tp - \tau) \vee \right) A_p \label{Diracflat} \ ,
\eea
where the tensor $\T$ and the quantity $\tau$ are defined around \eqref{TQ}, the dot in the derivatives indicates the action only on the form coefficient in flat indices, and the fluxes act with wedges and contractions in flat indices as (see also appendix \ref{ap:conv})
\bea
& H \w A_p = \frac{1}{3!}H_{abc}\, e^a\!\w e^b\!\w e^c\!\w A_p\ , \quad R\vee A_p = \frac{1}{3!} R^{abc} \, \iota_a \ \iota_b \ \iota_c A_p \ ,\label{fluxcontract}\\
& f \diamond A_p \equiv \frac{1}{2} f^{c}{}_{ab}\, \te^{a}\!\w \te^{b}\!\w \iota_c A_p\ ,\quad Q \diamond A_p \equiv \frac{1}{2} Q_{a}{}^{bc}\,\te^a\!\w \iota_b\ \iota_c A_p \ .\nn
\eea
Remarkably, this Dirac operator $\D$ is appearing in the pure spinors conditions, both for standard supergravity in \eqref{dP1stand} and \eqref{dP2stand}, and for $\b$-supergravity in \eqref{dP1intro} and \eqref{dP2intro}. A posteriori, it looks natural, since $\D$ acts on O(d,d) spinors; for Mink ($\mu=0$) with constant warp factor, \eqref{dP1stand} and \eqref{dP1intro} can thus be interpreted as (generalized) Dirac equations. Although this result was anticipated in \cite{Andriot:2014uda}, $\D$ was introduced there to study Bianchi identities (BI): $\D^2=0$ was shown to give BI of NSNS fluxes (together with a scalar condition) in absence of NS-branes.

We study further this Dirac operator and prove that it can be rewritten as
\bea
& \mbox{for standard supergravity:} \quad \D A_p = 2 e^{\p} e^{b\w} \d (e^{-b\w} e^{-\p} A_p)  \ , \label{Deb}\\
& \mbox{for $\b$-supergravity:} \quad \D A_p = 2 e^{\tp} e^{\b\vee} \d (e^{-\b\vee} e^{-\tp} A_p)  \ , \label{Debetaintro}
\eea
where $\b\vee\equiv \frac{1}{2} \b^{mn} \iota_m \iota_n = \frac{1}{2} \b^{ab} \iota_a \iota_b$. Although \eqref{Debetaintro} is a common guess, it has never been derived explicitly; we prove it here in appendix \ref{ap:betatwist}, obtaining first a helpful expression in curved indices \eqref{Dcurved}. In \eqref{Debetaintro} as well as in the generalized vielbein $\teee$ \eqref{genvielb}, $\b$ plays a role completely analogous to the $b$-field in \eqref{Deb} and $\eee$. The $b$-field there is responsible for a twist from $T\mmm \oplus T^*\mmm$ to the generalized tangent bundle $E_T$ of GCG, which can be viewed as a gerbe \cite{Belov:2007qj}. The $b$-field is then subject to the cocycle conditions (see e.g. \cite{Coimbra:2011nw}), that can be understood as global patching conditions, including $b$-field gauge transformations. We thus suggest for $\b$-supergravity a "$\b$-twist" of the local $T\mmm \oplus T^*\mmm$ into the generalized cotangent bundle $E_{T^*}$ discussed in \cite{Andriot:2013xca, Andriot:2014uda}. Obtaining the analogous to the cocycle conditions is the next step. We believe that this would be equivalent to characterising globally well-defined, i.e. geometric, backgrounds of $\b$-supergravity, a point discussed in details in \cite{Andriot:2014uda}.\footnote{The GCG (or analogous) formulation of $\b$-supergravity developed here, in particular the pure spinors conditions and related geometrical characterisation of vacua, should for this reason help getting a better handle on this ten-dimensional theory.} The equations derived in this paper should be considered on such backgrounds, otherwise just locally. A class of such backgrounds has been identified in \cite{Andriot:2014uda}: those admit isometries, and fields patch by $\b$-transforms (constant shifts of $\b$ along the isometries directions) and diffeomorphisms. Three explicit examples in this class are known: the toroidal example, the exotic $5_2^2$- or Q-brane \cite{deBoer:2010ud, Bergshoeff:2011se, deBoer:2012ma, Hassler:2013wsa, Geissbuhler:2013uka, Kimura:2014wga, Okada:2014wma, Kimura:2014bea}, and the dynamical SU(2) structure solution of \cite{McOrist:2010jw}.\footnote{\label{foot:SU2dyn}A warped Mink non-geometric background of type IIB supergravity with dynamical SU(2) structure is presented in \cite{McOrist:2010jw}. It is obtained from M-theory through type I by a chain of dualities. The internal space is locally a $T^4/\mathbb{Z}^2$ with complex coordinates $z_1$ and $w$, fibered over a $T^2$ along $z_2$. D7 or O7 are along the fiber, whose directions are isometries. Performing the field redefinition \eqref{fieldredef} towards $\b$-supergravity, one gets
\beq
\d \tilde{s}^2= e^{-\frac{3\varphi}{4}} \d s_{{\rm Mink}}^2 + e^{-\frac{3\varphi}{4}} (|\d w|^2 + |\d z_1|^2 ) + e^{\frac{3\varphi}{4}} |\d z_2|^2 \ , \ e^{\tp}=e^{-\frac{3\varphi}{2}} \ ,\ \b^{z_1 w}=2a\, z_2\ ,\ \b^{\ov{z_1} \ov{w}}=2a\, \ov{z_2} \ ,
\eeq
with $a$ a real constant and $\varphi(|z_2|)$. $\b$ patches with a $\b$-transform in O(4,4), while the metric is that of D7 or O7 (provided the warp factor is the good one), so it patches accordingly. Note that for a constant $\varphi$, this vacuum is the complex version of the toroidal example.} The question is whether geometric backgrounds of $\b$-supergravity exist beyond this class. We come back to these ideas in section \ref{sec:geomcharac}.

Finally, we further use the pure spinors conditions and the Dirac operator by looking at the superpotential $W$. Inspired by the literature on superpotentials for ${\cal N}=1$ four-dimensional effective theories obtained from ten-dimensional standard supergravities, in presence of an SU(3)$\times$SU(3) structure, we propose (considering only the NSNS contribution)
\beq
\tilde{W}_{{\rm NS}} = \frac{C}{2} \int_{\mmm} \langle e^{-\tp}\P_1^0 , \D \im \P_2^0 \rangle \ , \label{Wgenintro}
\eeq
for a constant warp factor. $C$ is a constant, the Mukai product is defined in \eqref{Mukai}, and for an SU(3) structure, the pure spinors are taken in the simple form
\beq
\Phi_+^0 = e^{\i \theta_+} e^{-iJ} \ , \qquad \Phi_-^0 = \i \Omega \ .\label{purespinSU3simpleintro}
\eeq
The formula \eqref{Wgenintro} reproduces standard supergravities superpotentials, choosing the Dirac operator \eqref{Diracstandard} and the dilaton $\p$ instead of $\tp$. This proposal is then expected to give the expressions to be obtained from $\b$-supergravity. For the SU(3) structure \eqref{purespinSU3simpleintro} and the Dirac operator \eqref{Diracflat}, we check, provided a few more assumptions, that \eqref{Wgenintro} reproduces formulas in the literature containing non-geometric fluxes. We get a good agreement with expressions of \cite{Aldazabal:2006up, Ihl:2007ah, Blumenhagen:2013hva} in type IIA and IIB, corresponding to an O6-plane and an O3- or O7-plane, while we obtain a new expression in the O5- or O9-plane (or heterotic) case. More details and references, as well as a discussion, can be found in section \ref{sec:W}.

The paper is organised as follows. The fermionic SUSY variations are obtained in section \ref{sec:fermsusy}. The compactification ansatz and consequent SUSY conditions are presented in section \ref{sec:compactsusy}, together with appendix \ref{ap:compact}. The pure spinors conditions are derived and discussed in section \ref{sec:purespin} and appendices \ref{ap:purespin} and \ref{ap:suff}. The work on the superpotential is presented in section \ref{sec:W}. The $\b$-twist and related geometrical characterisation are discussed in section \ref{sec:geomcharac} and appendices \ref{ap:betatwist} and \ref{ap:invD}. Finally, the $\b$-twist leads to further ideas: RR fluxes for $\b$-supergravity are discussed in section \ref{sec:RR}, while a possible relation to intermediate and dynamical SU(2) structure solutions is presented in section \ref{sec:dynSU2}. Further directions to explore are mentioned in section \ref{sec:outlook}. Conventions and notations are detailed in appendix \ref{ap:conv}.

\section{From the supersymmetry variations to the pure spinors conditions}

\subsection{Fermionic supersymmetry variations}\label{sec:fermsusy}

We explained in the Introduction that $\b$-supergravity is, in its current state, only formulated in its NSNS (bosonic) sector. The field redefinition relating its fields to those of standard supergravity is given by \cite{Andriot:2011uh} (see \cite{Andriot:2013xca, Andriot:2014uda} for more details)
\bea
& (g+b)^{-1}=(\tg^{-1}+\b) \ , \quad e^{-2 \tp} \sqrt{|\tg|} = e^{-2 \p} \sqrt{|g|} \ ,\label{fieldredef}\\
& \hhh= \eee^T \ \mathbb{I} \ \eee = \teee^T \ \mathbb{I} \ \teee \ ,\quad \eee= \begin{pmatrix} e & 0 \\ e^{-T} b & e^{-T} \end{pmatrix} \ , \ \teee= \begin{pmatrix} \te & \te \b \\ 0 & \te^{-T} \end{pmatrix} \ . \label{genvielb}
\eea
Since $\b$-supergravity is a local reformulation of standard supergravity (for instance, the Lagrangians of the two theories only differ by a total derivative), the existence of its supersymmetric completion is expected. We nevertheless do not have an explicit formulation for the latter, so we obtain here in an indirect way its fermionic SUSY variations. Some structures appearing in the Generalized Geometry formalism and in Double Field Theory (DFT), namely the Spin(9,1)$\times$Spin(1,9) derivatives \cite{Siegel:1993xq, Siegel:1993th, Hohm:2010xe, Jeon:2011cn, Coimbra:2011nw, Hohm:2011nu} or generalizations, were noticed to give these variations for standard supergravities (type II, heterotic and M-theory) \cite{Coimbra:2011nw, Hohm:2011nu, Jeon:2011sq, Jeon:2012hp, deWit:1986mz, Coimbra:2012af, Berman:2013cli, Godazgar:2014nqa, Coimbra:2014qaa}. These derivatives enter further quantities of supergravity (the Lagrangian, the equations of motion), and we have shown for the latter that the corresponding derivatives in $\b$-supergravity played exactly the same role there \cite{Andriot:2013xca, Andriot:2014uda}. Thus, as argued in the Introduction, we make the natural assumption that the derivatives give analogously in $\b$-supergravity the SUSY variations; this is what we now detail.

Type IIA and IIB standard supergravities have two pairs of chiral fermions; the NSNS contribution to their SUSY variations is given by
\bea
\delta \psi^{1,2}_{M}=& e^A{}_M \left(\N_{A}\mp \frac{1}{8} H_{ABC}\G^{BC}\right) \eps^{1,2} \ ,\\
\delta \rho^{1,2}=&\G^{A}\left(\N_{A}\mp \frac{1}{24} H_{ABC}\G^{BC}-\del_{A}\p\right) \eps^{1,2} \ , \nn
\eea
where notations are defined in the Introduction and appendix \ref{ap:conv}, and $\eps^{1,2}$ are the SUSY fermionic parameters, while the upper, lower, sign refers respectively to the number 1, 2. These conventions match those of \cite{Coimbra:2011nw}, except for the $\pm$ there denoted $1,2$ here, and the use here of flat indices. The above variations can be rephrased in terms of the following Spin(9,1)$\times$Spin(1,9) derivatives
\bea
D_{A}\eps^2=&\left(\N_{A}+ \frac{1}{8} H_{A \ov{BC}}\G^{\ov{BC}}\right) \eps^2\ ,\\
D_{\ov{A}}\eps^1=&\left(\N_{\ov{A}}- \frac{1}{8} H_{\ov{A}BC}\G^{BC}\right) \eps^1\ ,\nn\\
\G^{A}D_A \eps^1=&\left(\G^{A} \N_{A}- \frac{1}{24} H_{ABC}\G^{ABC}- \G^{A} \del_{A}\p\right) \eps^1 \ ,\nn\\
\G^{\ov{A}}D_{\ov{A}} \eps^2=& \left(\G^{\ov{A}} \N_{\ov{A}}+ \frac{1}{24} H_{\ov{ABC}}\G^{\ov{ABC}}- \G^{\ov{A}}\del_{\ov{A}}\p\right) \eps^2\ ,\nn
\eea
where the indices ${}_{A,\ \ov{A}}$ and spinors $\eps^1, \ \eps^2$ are here related to each Spin group respectively. One has \cite{Coimbra:2011nw}
\bea
& \delta \psi^1_M= e^{\ov{A}}{}_M D_{\ov{A}}\eps^1 \ ,\ \delta \psi^2_M=e^{A}{}_M D_{A}\eps^2\ ,\label{fermvarSpin}\\
& \delta \rho^1= \G^{A}D_A \eps^1 \ ,\ \delta \rho^2=\G^{\ov{A}}D_{\ov{A}} \eps^2\ ,\nn
\eea
using \eqref{g2}, and considering the vielbeins aligned. As explained previously, we consider now the fermionic SUSY variations of $\b$-supergravity to be given analogously by \eqref{fermvarSpin}, where we replace the vielbein $e$ by $\te$, the fermions $\psi^{1,2}_M, \rho^{1,2}$ by $\tilde{\psi}^{1,2}_M, \tilde{\rho}^{1,2}$, and use the following derivatives determined in \cite{Andriot:2013xca}
\bea
D_{A} \eps^2 =& \left( \N_A - \eta_{AD} \cN^D - \frac{1}{8} \eta_{AD} \eta_{\ov{BE}}\eta_{ \ov{CF}} R^{D\ov{EF}} \G^{\ov{BC}} \right) \eps^2\ ,\label{Derbetasugra}\\
D_{\ov{A}} \eps^1 =& \left( \N_{\ov{A}} + \eta_{\ov{AD}} \cN^{\ov{D}} - \frac{1}{8} \eta_{\ov{AD}} \eta_{BE} \eta_{CF} R^{\ov{D}EF} \G^{BC} \right) \eps^1 \ , \nn\\
\G^A D_A \eps^1 = & \left( \G^A \N_A - \G^A \eta_{AD} \cN^D  + \frac{1}{24} \eta_{AD} \eta_{BE} \eta_{CF} R^{DEF} \G^{ABC} - \G^A \del_A \tp - \G^A \eta_{AB} (\b^{BC} \del_C \tp - \T^B) \right) \eps^1 \ ,\nn\\
\G^{\ov{A}} D_{\ov{A}} \eps^2 = & \left( \G^{\ov{A}} \N_{\ov{A}} + \G^{\ov{A}} \eta_{\ov{AD}} \cN^{\ov{D}}  + \frac{1}{24} \eta_{\ov{AD}} \eta_{\ov{BE}} \eta_{\ov{CF}} R^{\ov{DEF}} \G^{\ov{ABC}} - \G^{\ov{A}} \del_{\ov{A}} \tp + \G^{\ov{A}} \eta_{\ov{AB}} (\b^{\ov{BC}} \del_{\ov{C}} \tp - \T^{\ov{B}}) \right) \eps^2 \ ,\nn
\eea
where the covariant derivatives $\N_A$ and $\cN^A$ are defined in appendix \ref{ap:conv}, and on spinors as
\bea
\N_A \eps & = \del_A \eps + \frac{1}{4} \o_A{}^B{}_C \eta_{BD} \G^{DC} \eps \ , \\
\cN^A \eps & = - \b^{AB}\del_B \eps + \frac{1}{4} \co^A{}_C{}^B \eta_{BD} \G^{DC} \eps \ ,
\eea
and $\co$ is the spin connection of $\cN$, related to the $Q$-flux as in \eqref{coQ}, that was denoted $\o_Q$ in \cite{Andriot:2013xca, Andriot:2014uda}. The tensor $\T^A=\N_B \b^{AB}$ often appears with the dilaton in $\b$-supergravity, because the above combination comes from the $\mathbb{R}^+$ factor in the Generalized Geometry formalism \cite{Andriot:2013xca, Andriot:2014uda}; later on we will use the following expression and notation
\beq
\T^A=-Q_B{}^{BA} + \frac{1}{2} \b^{BC} f^A{}_{BC} \ ,\quad \tau^A\equiv - \frac{1}{2}\b^{BC}f^{A}{}_{BC} \ .  \label{TQ}
\eeq
We refer to the Introduction and appendix \ref{ap:conv} for more conventions. From \eqref{fermvarSpin} and \eqref{Derbetasugra}, with aligned vielbeins, we deduce the NSNS contribution to the fermionic SUSY variations of both a type IIA and IIB $\b$-supergravity: it is given by \eqref{fermvarintro}, that we repeat here for convenience
\bea
\delta \tilde{\psi}^{1,2}_{M}=& \te^A{}_M \left( \N_{A} \pm \eta_{AD} \cN^{D} - \frac{1}{8} \eta_{AD} \eta_{BE} \eta_{CF} R^{DEF} \G^{BC} \right) \eps^{1,2} \label{fermvar}\\
\delta \tilde{\rho}^{1,2}=&\left( \G^A \N_A \mp \G^A \eta_{AD} \cN^D  + \frac{1}{24} \eta_{AD} \eta_{BE} \eta_{CF} R^{DEF} \G^{ABC} - \G^A \del_A \tp \mp \G^A \eta_{AB} (\b^{BC} \del_C \tp - \T^B) \right) \eps^{1,2} \ .\nn
\eea

\subsection{Compactification ansatz and resulting conditions for a supersymmetric vacuum}\label{sec:compactsusy}

We now specify an ansatz for the fields, suited to the compactification of a ten-dimensional background on a compact internal six-dimensional manifold $\mmm$. Then, we will study the decomposition of the previous SUSY variations accordingly, and deduce conditions for a SUSY vacuum. To start with, we consider the following ten-dimensional metric
\beq
\d \tilde{s}^2 = e^{2A(y)} \tg_{\mu\nu} (x) \d x^\mu \d x^\nu + \tg_{mn} (y) \d y^m \d y^n \ ,
\eeq
where Latin indices are the internal ones, and Greek indices are the four-dimensional ones; $e^{2A}$ is the warp factor. The ten-dimensional vielbeins are then decomposed into
\beq
\te^{A=\alpha}{}_{M=\mu}= \te^{\al}{}_{\mu}=e^{A(y)}\te^{\dot{\al}}{}_{\mu} (x)\ ,\quad \te^{A=a}{}_{M=m}=\te^{a}{}_{m} (y) \ .
\eeq
The ten-dimensional $\b^{AB}$ is chosen a priori non-trivial only along the internal directions, i.e.
\beq
\b^{\al\b}=\b^{\al b}=\b^{a \b}=0\ ,\ \b^{ab}(y) \ .
\eeq
This ansatz allows to compute the various components of the ten-dimensional fluxes $f,\ Q,\ R$, and those of the spin connections $\omega$ and $\co$. This is detailed in appendix \ref{ap:compact}.

We now turn to the spinors, in particular to the ten-dimensional SUSY parameters. In agreement with the above metric, the Lorentz group and its spinorial representation is split in two factors; $\eps^{1,2}$ should accordingly be decomposed on a basis of internal spinors. However, we restrict ourselves to backgrounds with ${\cal N}=1$ preserved SUSY in four dimensions, so consider only one external spinor $\zeta_+$. We are then left with two internal spinors $\eta^{1,2}_+$, and the decomposition is
\beq
\mbox{IIA}\ \begin{cases} \eps^1 = \zeta_+ \otimes \eta^{1}_+ + \zeta_- \otimes \eta^{1}_- \\ \eps^2 = \zeta_+ \otimes \eta^{2}_- + \zeta_- \otimes \eta^{2}_+ \end{cases} \ ,\qquad \mbox{IIB}\ \begin{cases} \eps^1 = \zeta_+ \otimes \eta^{1}_+ + \zeta_- \otimes \eta^{1}_- \\ \eps^2 = \zeta_+ \otimes \eta^{2}_+ + \zeta_- \otimes \eta^{2}_- \end{cases} \ ,
\eeq
the distinction between the two theories coming from the chiralities denoted with $\pm$. The six-dimensional and four-dimensional spinors are Weyl, and Euclidian respectively Lorentzian. This implies the complex conjugations $(\eta^i_+)^*=\eta^i_-$ and $(\zeta_+)^*=\zeta_-$. We also consider $\eta_+^i(y)$ and $\zeta_+(x)$. The ten-dimensional spinors are Majorana-Weyl, hence real, as given by the sum of the two terms in $\eps^i$. The ten-dimensional $\G$-matrices are decomposed similarly, in terms of the six-dimensional $\g^a$ and four-dimensional $\g^{\al}$. The properties of the latter, together with the components of the spin connections, allow to obtain the various components of the two spinorial covariant derivatives $\N_A$ and $\cN^A$. This is detailed in appendix \ref{ap:compact}.

Finally, let us further specify the external part: the four-dimensional space-time is chosen maximally symmetric. The three possibilities (Mink, AdS and de Sitter) forbid to single out a vector, implying $\del_{\al} \tp=0$ in the background. In addition, the four-dimensional covariant derivative $\N_{\al} \zeta_\pm$, generically decomposed on a spinor basis, is then at most given by
\beq
\N_{\al} \zeta_\pm =\frac{1}{2} \mu_{\pm} e^{-A} \eta_{\al \b} \g^{\b} \zeta_\mp\ . \label{4dNzeta}
\eeq
Let us comment on the coefficient. First, one can verify that $\N_{\al} $ carries a factor $e^{-A}$. Indeed, starting with the coordinate derivative $\del_{\mu}$, the multiplication by $\te^{\mu}{}_{\al}$ brings such a factor; also, one has $\o_{\al}{}^{\b}{}_{\g}= e^{-A} \o_{\dot{\al}}{}^{\dot{\b}}{}_{\dot{\g}}$. The matrices $\g^{\al}$ do not carry such a factor, since they satisfy the Clifford algebra. So we get $\N_{\al} = e^{-A} \N_{\dot{\al}} $. This factor, manifest on the RHS of \eqref{4dNzeta}, then carries the whole dependence on the internal coordinates. The generic $\mu_{\pm}$ is thus restricted to be a complex function of the external coordinates.\footnote{\label{foot:signeps}$\mu_{\pm}$ defined in \eqref{4dNzeta} can a priori change according to the theory IIA or IIB, although we do not denote it differently here. In \cite{Koerber:2007xk}, this quantity is related to the vacuum value of the superpotential, with a sign change in between the theories. Here we will rather introduce the sign $\varepsilon$.} For Mink and AdS (de Sitter does not allow to consider such a spinorial equation), the value $|\mu_{\pm}|^2$ is actually known to be related to the scalar curvature, i.e. to the cosmological constant.\footnote{Using relations with superpotential and scalar potential, the equality $3|\mu_-|^2= - \Lambda$ was derived in \cite{Koerber:2007xk}. We give here an alternative derivation. We use dotted flat indices for a purely external dependence; $\eta_{\dot{\al} \dot{\b}}$ and $\gamma^{\dot{\al}}$ are the same as without dot, and are constant. On the one hand, from $4 [\N_{\dot{\al}}, \N_{\dot{\b}} ] \zeta_+= \R_{\dot{\al}\dot{\b}\dot{\gamma}\dot{\delta}} \gamma^{\dot{\gamma}\dot{\delta}} \zeta_+ $, one gets using \eqref{g2} and symmetry properties $2 \gamma^{\dot{\b}} [\N_{\dot{\al}}, \N_{\dot{\b}} ] \zeta_+= - \R_{\dot{\al}\dot{\b}} \gamma^{\dot{\b}} \zeta_+ $. On the other hand, using definitions of $\N$, symmetry properties of $\omega$, and $[\gamma^{\dot{\al}}, \gamma^{\dot{\b}\dot{\gamma}} ]=4 \eta^{\dot{\al} [\dot{\b}} \gamma^{\dot{\gamma}]} $, one shows $\N_{\dot{\al}}(\gamma^{\dot{\b}} \zeta_{\pm})=\gamma^{\dot{\b}} \N_{\dot{\al}}( \zeta_{\pm})$. From this, \eqref{4dNzeta} and \eqref{g2}, one gets $2 \gamma^{\dot{\b}} [\N_{\dot{\al}}, \N_{\dot{\b}} ] \zeta_+= 3 \mu_+ \mu_- \eta_{\dot{\al}\dot{\b}} \gamma^{\dot{\b}} \zeta_+ $. With $\R_4= 4 \Lambda$, we conclude $3 \mu_+ \mu_- = - \Lambda$.} We thus restrict $\mu_{\pm}$ to be complex constants, differing at most by a phase. SUSY will impose $\mu_+$ to be the complex conjugate of $\mu_-\equiv \mu$.\\

We are now interested in vacua satisfying the compactification ansatz just described and preserving $\mathcal N=1$ SUSY. While not strictly necessary for SUSY, we make the additional helpful assumption that the internal spinors $\eta^i_+$ are globally defined and non-vanishing. Given a metric and an orientation on $\mmm$, this further assumption reduces the structure group of the tangent bundle, for one spinor to SU(3), and for two (i.e. not parallel) to SU(2). Requiring the existence of such spinor(s), i.e. having a reduced structure group, is a useful topological constraint allowing to consider equivalently specific globally defined forms on $\mmm$. In GCG, the structure group of $T\mmm \oplus T^*\mmm$ is reduced to SU(3)$\times$SU(3) by the existence of globally defined $\eta^i_+$. This equivalently defines the pure spinors $\P_{\pm}$ that can be viewed as polyforms (sums of forms of different degrees), as described in section \ref{sec:purespin}. In addition to this topological condition, the spinors will have to satisfy differential conditions. These are derived from the fermionic SUSY variations, that have to vanish in a SUSY background. We obtain these conditions in the following, by setting \eqref{fermvar} to zero and imposing in these equations the above compactification ansatz (using as well material of appendix \ref{ap:compact}).

We start with the vanishing variation of the gravitini \eqref{fermvar}: in type IIB, it gives on the internal directions
\bea
& \zeta_+ \otimes \left( \N_{a} \pm \eta_{ad} \cN^{d} - \frac{1}{8} \eta_{ad} \eta_{be} \eta_{cf} R^{def} \g^{bc} \right) \eta^{1,2}_+\\
+& \zeta_- \otimes \left( \N_{a} \pm \eta_{ad} \cN^{d} - \frac{1}{8} \eta_{ad} \eta_{be} \eta_{cf} R^{def} \g^{bc} \right) \eta^{1,2}_- = 0 \ ,\nn
\eea
and in type IIA one should change the chirality on the $\eta^2$. Projecting by chirality imposes both lines to vanish. The two lines are also complex conjugate, so the only conditions are (in type IIB)
\beq
\left( \N_{a} \pm \eta_{ad} \cN^{d} - \frac{1}{8} \eta_{ad} \eta_{be} \eta_{cf} R^{def} \g^{bc} \right) \eta^{1,2}_+ =0 \ .
\eeq
On the external directions, we obtain in type IIB
\bea
& \left( \N_{\al} \otimes \id  +  \frac{1}{2} \eta_{\al \b} \g^\b \g_{(4)}\otimes \g^d \del_d A \pm \frac{1}{2} \eta_{\al \delta} \g^{\delta} \g_{(4)}\otimes \g^c \eta_{c d} \b^{de} \del_{e} A \right)\eps^{1,2} =0 \\
\Leftrightarrow \quad &  \frac{1}{2} \eta_{\al \b} \g^\b \zeta_+ \otimes \left( \mu_{-} e^{-A} \eta_-^{1,2} + (\g^d \del_d A \pm  \g^c \eta_{c d} \b^{de} \del_{e} A) \eta_+^{1,2} \right)  \nn\\
+ & \frac{1}{2} \eta_{\al \b} \g^\b \zeta_- \otimes \left( \mu_{+} e^{-A} \eta_+^{1,2} -(\g^d \del_d A \pm \g^c \eta_{c d} \b^{de} \del_{e} A) \eta_-^{1,2} \right) =0 \ ,\nn
\eea
and for type IIA one should change the chirality of the $\eta^2$. Again, both lines should vanish, from which we deduce $\mu_+^*=\mu_-\equiv \mu$. So the only conditions are (in type IIB)
\beq
\mu\ \eta_-^{1,2} + e^{A} \g^d \left( \del_d A \pm  \eta_{dc} \b^{ce} \del_{e} A \right) \eta_+^{1,2} = 0 \ .
\eeq
Note that having Mink ($\mu=0$) is equivalent to a constant warp factor; this is due to the fact we only have NSNS contributions.\footnote{We reached the same conclusion in \cite{Andriot:2013xca}, using the equations of motion and a few more assumptions, such as a constant dilaton. Here, it comes from SUSY.} We finally turn to the variation $\delta \tilde{\rho}^{1,2}$ in \eqref{fermvar}. A few computations, using in particular $\g^{\al} \eta_{\al \b} \g^{\b}=4$, lead to
\bea
\delta \tilde{\rho}^{1,2} =& \Bigg(\g^{\al} \N_{\al} \otimes \id + \g_{(4)} \otimes \frac{1}{24} \eta_{ad} \eta_{be} \eta_{cf} R^{def} \g^{abc} \\
& + \g_{(4)} \otimes \g^a \Big( \N_a + \del_a (2A-\tp) \mp \eta_{ad} \cN^d \mp  \eta_{ad} \b^{de} \del_e (2A + \tp) \pm  \eta_{ad} \T^d \Big) \Bigg) \eps^{1,2} \nn\\
&\!\!\!\!\!\!\!\!\!\!\!\!\!\!\!\!\!\!\!\!\!\!\!\!\!\!\! = \zeta_+ \otimes \left( 2 \mu_- e^{-A} \eta_-^{1,2} + \Big( \frac{1}{24} \eta_{ad} \eta_{be} \eta_{cf} R^{def} \g^{abc} + \g^a \Big( \N_a + \del_a (2A-\tp) \mp \eta_{ad} \cN^d \mp  \eta_{ad} \b^{de} \del_e (2A + \tp) \pm  \eta_{ad} \T^d \Big)\Big) \eta_+^{1,2}  \right)  \nn\\
&\!\!\!\!\!\!\!\!\!\!\!\!\!\!\!\!\!\!\!\!\!\!\!\!\!\!\! + \zeta_- \otimes \left( 2 \mu_+ e^{-A} \eta_+^{1,2} - \Big( \frac{1}{24} \eta_{ad} \eta_{be} \eta_{cf} R^{def} \g^{abc} + \g^a \Big( \N_a + \del_a (2A-\tp) \mp \eta_{ad} \cN^d \mp  \eta_{ad} \b^{de} \del_e (2A + \tp) \pm  \eta_{ad} \T^d \Big)\Big) \eta_-^{1,2}  \right) \nn
\eea
for type IIB, while for type IIA one should change the chirality on the $\eta^2$. Setting this variation to zero imposes both lines to vanish, from which we deduce again $\mu_+^*=\mu_-=\mu$, and the two lines are then complex conjugate. We are left with
\beq
2 \mu \ \eta_-^{1,2} + e^{A} \Big( \frac{1}{24} \eta_{ad} \eta_{be} \eta_{cf} R^{def} \g^{abc} + \g^a \Big( \N_a + \del_a (2A-\tp) \mp \eta_{ad} \cN^d \mp  \eta_{ad} \b^{de} \del_e (2A + \tp) \pm  \eta_{ad} \T^d \Big)\Big) \eta_+^{1,2} = 0 \ .\nn
\eeq
To summarize, the backgrounds of interest satisfy the above compactification ansatz and admit an SU(3)$\times$SU(3) structure. In addition, they verify in type IIB the following three constraints, namely the SUSY conditions or Killing spinor equations
\bea
& \mu\ \eta_-^{1,2} + e^{A} \left( \s{\del} A \pm  \s{\b}_{\del} A \right) \eta_+^{1,2} = 0 \label{susy1}\\
&  \N_{a}\eta^{1,2}_+ = \left(\mp \eta_{ad} \cN^{d} + \frac{1}{8} \eta_{ad} \eta_{be} \eta_{cf} R^{def} \g^{bc} \right) \eta^{1,2}_+ \label{susy2}\\
& \s{\N}\eta_+^{1,2} = - 2 \mu e^{-A} \eta_-^{1,2} - \Big( \frac{1}{4} \s{R} +  \s{\del} (2A-\tp) \mp \s{\cN} \mp  \s{\b}_{\del} (2A + \tp) \pm  \s{\T} \Big) \eta_+^{1,2} \ ,\label{susy3}
\eea
(for type IIA one should change the chirality on the $\eta^2$) where we introduced the notations
\beq
\s{\del} = \g^a \del_a \ ,\ \s{\N}=\g^a \N_a \ ,\ \s{\b}_{\del}= \g^a \eta_{ab} \b^{bc} \del_{c}\ ,\ \s{\cN}= \g^a\eta_{ab}\cN^b \ ,\ \s{\T}=\g^a \eta_{ab} \T^b \ ,\  \s{R}=\frac{1}{6} \eta_{ad} \eta_{be} \eta_{cf} R^{def} \g^{abc} \ . \nn
\eeq

\subsection{Supersymmetry conditions in terms of pure spinors}\label{sec:purespin}

We now want to formulate the previous SUSY conditions using pure spinors $\P_{\pm}$, for motivations discussed in the Introduction. To do so, we follow closely the procedure described in the appendix of \cite{Grana:2006kf}. $\P_{\pm}$ are defined at first as the following bispinors
\beq
\P_+ = \eta^{1}_+ \otimes \eta^{2 \dag}_+,\quad \P_- = \eta^{1}_+ \otimes \eta^{2 \dag}_- \ . \label{purespinordef}
\eeq
The product can be expressed thanks to the Fierz identity given in six dimensions by
\beq
\eta^{1}_+ \otimes \eta^{2\dag}_\pm=\frac{1}{8}\sum_{k=0}^6 \frac{1}{k!}\left( \eta^{2\dag}_\pm \gamma_{a_k \dots a_1} \eta^{1}_+ \right ) \gamma^{a_1 \dots a_k} \ ,
\eeq
where indices are lowered by the flat metric. In addition, the Clifford map relates antisymmetric products of $\g$-matrices and differential forms
\beq
C=\sum_k \frac{1}{k!} C^{(k)}_{a_1 \dots a_k} \te^{a_1}\w \ldots \w \te^{a_k} \ \longleftrightarrow \slashed{C}=\sum_k \frac{1}{k!} C^{(k)}_{a_1 \dots a_k} \g^{a_1 \dots a_k} \ . \label{cliffmap}
\eeq
Thanks to this map, the above pure spinors can be viewed as polyforms, i.e. sums of forms of different degrees; note that $\P_{\pm}$ in \eqref{purespinordef}, expressed with the Fierz identity, should be understood as slashed. $\P_{\pm}$ are examples of Spin(6,6) spinors on $T\mmm \oplus T^*\mmm$, as considered in GCG. Being bispinors, they are as well Spin(6)$\times$Spin(6) spinors; similarly, they are pure because they are built from two pure spinors (any spinor is pure in six dimensions). As mentioned in the previous section, we require them to be globally defined, which reduces the structure group of $T\mmm \oplus T^*\mmm$ to SU(3)$\times$SU(3). On the tangent bundle, this gets declined into an SU(3) or an SU(2) structure group. For example, $\P_{\pm}$ for an SU(3) structure are given as polyforms in \eqref{purespinSU3simple} in a simplified case. More details on such examples are provided in section \ref{sec:purespinWstand}. The pure spinors are acted on by Cliff(6,6) $\G$-matrices, from which one can construct a chirality operator. Through the Clifford map, their chirality is simply related to the degree of the forms, i.e. the summation runs only over forms of even, respectively odd, degree, for positive, respectively negative, chirality. This is equivalent to the number of $\g$-matrices, so via the Fierz identity, it is related to the chiralities of the $\eta^i$: $\P_{+}$ or $\P_{-}$ is of positive or negative chirality.

Reformulating the SUSY conditions on the $\eta^i$ \eqref{susy1} - \eqref{susy3} as polyform equations on the pure spinors essentially amounts to compute the exterior derivative
\bea
\d \P_{\pm} \equiv \te^a \w \N_a \P_{\pm} \ . \label{diffspinor}
\eea
To do so, we write \eqref{diffspinor} with $\g$-matrices acting on the $\eta^i$, thanks to the Clifford map and the bispinor expressions. We then use the above SUSY conditions, and finally rewrite the resulting expression in terms of forms, using the Clifford map backwards. The whole procedure, with the required properties of the Clifford map, are detailed in appendix \ref{ap:purespin}. Note the following subtlety. We obtain at first expressions for $\d \P_{\pm}$ that are simpler than the final ones \eqref{dP1intro} and \eqref{dP2intro}. Establishing them only required to use \eqref{susy2} and \eqref{susy3}, but not \eqref{susy1}. This is due to the absence of a RR contribution. We nevertheless follow further the procedure of \cite{Grana:2006kf} for standard supergravity, and construct from \eqref{susy1} another form expression, that should be given by the RR fluxes but is here vanishing. We add this quantity (as in \cite{Grana:2006kf}) to one of the $\d \P_{\pm}$ obtained. This eventually results in \eqref{dP1intro} and \eqref{dP2intro}, that we repeat here for convenience
\bea
\!\!\!\!\!\!\!\!\! e^{\tp}\left(\d -  \cN^a \cdot \iota_a+\T \vee + R \vee\right) \left(e^{-\tp} \P_1\right) + e^{-2A}& \left( \d  + \cN^a \cdot \iota_a \right)(e^{2A}) \P_1 =2\varepsilon\ e^{-A} \mu\ \re(\Phi_2) \label{dP1}\\
\!\!\!\!\!\!\!\!\! e^{\tp}\left(\d -  \cN^a \cdot \iota_a+\T \vee + R \vee\right) \left(e^{-\tp} \P_2\right) + e^{-2A}& \left( \d  + \cN^a \cdot \iota_a \right)(e^{2A}) \P_2 \label{dP2} \\
\!\!\!\!\!\!\!\!\! = 3 \varepsilon\ e^{-A} &\ \i \im(\overline{\mu} \Phi_1) + e^{-A}\left( \d  - \cN^a \cdot \iota_a \right)(e^{A}) \ov{\P_2} \ , \nn
\eea
where $\P_{1,2}$ and $\varepsilon$ depend on the theory \eqref{IIAB}. The sign $\varepsilon$ can be viewed as a change of $\mu$ in between the two theories, see footnote \ref{foot:signeps}.\\

Let us comment on these pure spinors conditions, and compare them to those of standard supergravity given in \eqref{dP1stand} and \eqref{dP2stand}. The NSNS sector of the two theories are known to match for vanishing $b$ and $\b$. Here, one can verify that the pure spinors conditions do agree in that case, which is a non-trivial check of our result. More generally, it is remarkable (and another confirmation of our result) that the differential operator acting on the pure spinors in both theories is precisely the Dirac operator $\D$ discussed in the Introduction. This was anticipated in \cite{Andriot:2014uda} where $\D$ was computed for the two theories, as given in \eqref{Diracstandard} and \eqref{Diraccurved}. This Dirac operator $\D = \G^{{\cal A}} D_{{\cal A}}$ was obtained from the Spin(d,d)$\times \mathbb{R}^+$ derivative $D_{{\cal A}}$ for a $d$-dimensional space-time \cite{Grana:2008yw, Hohm:2011dv, Coimbra:2011nw, Jeon:2011vx, Jeon:2012kd, Jeon:2012hp, Geissbuhler:2013uka, Andriot:2013xca} and the Cliff(d,d) $\G^{{\cal A}}$-matrices, for which we took a representation as forms and contractions. Using the expressions of $\D$ in terms of fluxes \eqref{Diracflatstandard} and \eqref{Diracflat}, the nilpotency $\D^2=0$ was shown to give the Bianchi identities of the NSNS fluxes \cite{Andriot:2014uda}. Here, $\P_{1,2}$ being Spin(6,6) spinors, they are naturally acted on by the Dirac operator in both theories, and the pure spinors conditions can be viewed as Dirac equations (with RHS).

Let us now focus on the particular case of a Mink space-time ($\mu=0$): it provides an interesting characterisation of the background in standard supergravity, as discussed in the Introduction. Here, \eqref{dP1} and \eqref{dP2} reduce to
\bea
& e^{\tp}\left(\d -  \cN^a \cdot \iota_a+\T \vee + R \vee\right) \left(e^{-\tp} \P_1\right) + e^{-2A}\left( \d  + \cN^a \cdot \iota_a \right)(e^{2A}) \P_1 = 0 \label{dP1M}\\
& e^{\tp}\left(\d -  \cN^a \cdot \iota_a+\T \vee + R \vee\right) \left(e^{-\tp} \re \P_2\right) + e^{-A}\left( \d  + 3 \cN^a \cdot \iota_a \right)(e^{A}) \re \P_2 = 0  \label{dP2M1}\\
& e^{\tp}\left(\d -  \cN^a \cdot \iota_a+\T \vee + R \vee\right) \left(e^{-\tp} \im \P_2\right) + e^{-A}\left( 3 \d  + \cN^a \cdot \iota_a \right)(e^{A}) \im \P_2 = 0 \ . \label{dP2M2}
\eea
In contrast to standard supergravity, the warp factor terms can here not be factorised with the dilaton, because of the sign in front of $\cN^a (e^{2A}) \iota_a$. In other words, the condition \eqref{dP1M} cannot be written (in full generality) as a pure spinor closed under the Dirac operator $\D$ of \eqref{Diraccurved}. We do not get the analogous to the GCY condition, but still have the analogous to the generalized complex structure condition, as discussed in section \ref{sec:more}. Since these warp factor terms cannot be absorbed within $\D$, they can be understood as a (new) effect due to the compactification, that goes beyond the manifold $\mmm$ and the Spin(6,6)$\times \mathbb{R}^+$ structure of $\D$; they are reminiscent of the underlying ten dimensions. Nevertheless, in the particular case of backgrounds for which $\b^{ab}\del_b A=0$, e.g. when $\b$ is only non-zero along isometry directions (see \cite{Andriot:2014uda, Andriot:2011uh} for related discussions), the warp factor terms can be factorised. We are then back to a situation analogous to standard supergravity, where the pure spinors conditions are expressed purely in terms of $\D$ (and RR). The corresponding background characterisation is discussed in section \ref{sec:more}.\\

The pure spinors conditions \eqref{dP1} and \eqref{dP2} have been derived using the three SUSY conditions \eqref{susy1} - \eqref{susy3}, meaning that the former are necessary for SUSY to be preserved in the backgrounds considered. It is important to study whether they are also sufficient: this would guarantee that a solution to \eqref{dP1} and \eqref{dP2} does preserve SUSY. In addition, this would allow to trade solving Killing spinor equations for form equations, which is of practical interest. Following the method of \cite{Grana:2006kf}, we address this question in appendix \ref{ap:suff}. We conclude that \eqref{dP1} and \eqref{dP2} are not sufficient: they allow for a remaining freedom or ambiguity with respect to the SUSY conditions. However, we argue that this ambiguity should be fixed in presence of RR fluxes. Moreover, the RR contribution is expected to simply consist in an addition to the RHS of \eqref{dP2}, in analogy to standard supergravity with \eqref{dP2stand}. Therefore, the results established in this paper remain useful, the discussion on the structures appearing, such as the Dirac operator and the related background characterisation, is in any case relevant.

\section{The superpotential}\label{sec:W}

In this section, we first come back to pure spinors defining an SU(3)$\times$SU(3) structure and discuss their properties. They allow to write down an expression for the NSNS part of the superpotential obtained from standard supergravities, as reviewed. The appearance of the Dirac operator leads us to propose a corresponding expression from $\b$-supergravity, that includes non-geometric fluxes. This proposal is compared to the literature and further discussed.

\subsection{SU(3)$\times$SU(3) structure pure spinors and standard ${\cal N}=1$ superpotential}\label{sec:purespinWstand}

The pure spinors $\P_{\pm}$ have been defined as bispinors \eqref{purespinordef} in terms of the internal spinors $\eta_{\pm}^{1,2}$. For globally defined $\eta^i_+$, one gets an SU(3)$\times$SU(3) structure group on $T\mmm \oplus T^*\mmm$ (see sections \ref{sec:compactsusy} and \ref{sec:purespin}). The corresponding structure group of the tangent bundle then depends on the two internal spinors: for $\eta_{+}^{1}$ and $\eta_{+}^{2}$ parallel, i.e. proportional, one gets an SU(3) structure; otherwise one has an SU(2) structure. For the latter, the differential conditions on the pure spinors impose to distinguish further two cases: for $\eta_{+}^{1}$ and $\eta_{+}^{2}$ orthogonal, i.e. related by a gamma matrix, one gets an orthogonal (or static) SU(2) structure; if the two spinors are neither parallel nor orthogonal, one gets an intermediate SU(2) structure. If the angle between the spinors is actually varying on $\mmm$, one talks of a dynamical SU(2) structure. We discuss more and provide references on these various cases in section \ref{sec:dynSU2}.

Thanks to the Fierz identity and the Clifford map, the pure spinors can be viewed as polyforms. The corresponding formulas for $\Phi_{\pm}$ vary accordingly to the cases just mentioned: the different expressions can be found e.g. in \cite{Andriot:2010sya}. One has for an SU(3) structure
\beq
\Phi_+ = \frac{|a|^2}{8} e^{\i \theta_+} e^{-iJ} \ , \qquad \Phi_- = - \i e^{\i \theta_-} \frac{|a|^2 }{8}\Omega \ ,\label{purespinSU3}
\eeq
where $J$ is a real (1,1)-form and $\Omega$ is a (3,0)-form, with respect to an almost complex structure. These forms satisfy further conditions we will come back to; for a CY they are the K\"ahler form and the holomorphic $3$-form. $|a|$ is related to the norm of the internal spinors; those are taken here to be of the same norm, as is the case in presence of an orientifold plane. The latter in turn further sets $|a|^2=e^A$, that we will also use. In the following we will consider a constant warp factor and constant phases $\theta_{\pm}$, as done for most of the formulas for the superpotential in the literature. The orientifold when present then fixes the phase $\theta_+$, while $\theta_-$ is left free (that phase is not physical) \cite{Koerber:2007hd}. We thus choose for convenience $\theta_-=\pi$, while $\theta_+$ will be fixed as
\beq
\mbox{O3 or O7}: e^{\i\theta_+}=\pm \i \ ,\ \mbox{O5 or O9}: e^{\i\theta_+}=\pm 1 \ ,\ \mbox{O6}: e^{\i\theta_+}\ \mbox{is free.} \label{phasechoice}
\eeq
Note that O4- or O8-planes do not allow for an SU(3) structure. Also, for an O6-plane, $e^{\i\theta_+}$ is sometimes taken to be $1$ in the literature. Given the fixing of these various parameters, we will consider in the following the simpler SU(3) structure pure spinors \eqref{purespinSU3simpleintro}
\beq
\Phi_+^0 = e^{\i \theta_+} e^{-iJ} \ , \qquad \Phi_-^0 = \i \Omega \ .\label{purespinSU3simple}
\eeq
With the above assumptions, the pure spinors conditions for SUSY in standard supergravity \eqref{dP1stand} and \eqref{dP2stand} simplify to
\bea
& e^{\p}\left(\d -  H\w\right) \left(e^{-\p} \P_1^0\right) =2\varepsilon\ e^{-A} \mu\ \re(\Phi_2^0) \label{dP1stand0}\\
& e^{\p}\left(\d -  H\w\right) \left(e^{-\p} \P_2^0\right) =3\varepsilon\ e^{-A}\ \i \im(\overline{\mu} \Phi_1^0) +\ \mbox{RR} \ . \label{dP2stand0}
\eea

In general, two pure spinors $\Phi_1$ and $\Phi_2$ of GCG defining an SU(3)$\times$SU(3) structure satisfy some compatibility conditions. Those include conditions on the norms, that need not be specified here, and the following
\beq
\langle \P_1, (v \vee + \xi \w) \P_2 \rangle = \langle \ov{\P_1}, (v \vee + \xi \w) \P_2 \rangle  =0\ , \quad \forall (v+\xi) \in T\mmm\oplus T^*\mmm \ , \label{compat0}
\eeq
where the Mukai product is defined as taking the top (here six) form
\beq
\langle \Psi_1 , \Psi_2 \rangle = \Psi_1 \w \lambda( \Psi_2 )|_{\mbox{top}} \ , \label{Mukai}
\eeq
and $\lambda$ brings a sign by reversing all the form indices. The compatibility condition \eqref{compat0} can also be formulated with matrices $\Gamma^{\cal A}$. In the case of an SU(3) structure, this leads to the condition
\beq
J \w \Omega= 0 \label{compat} \ ,
\eeq
which can also be understood from the almost complex structure.\\

We now turn to the superpotential for which we will use the various properties just described. The superpotential $W$ of the ${\cal N}=1$ four-dimensional effective theory obtained from standard ten-dimensional supergravities has been formulated in terms GCG pure spinors: this was done for an SU(3) structure \cite{Grana:2005ny, Benmachiche:2006df} (see also \cite{Villadoro:2005cu}), and then for an SU(3)$\times$SU(3) \cite{Grana:2006hr, Koerber:2007xk, Cassani:2007pq} (see also \cite{Koerber:2010bx} for a uniform presentation; note the last three references contain warp factors). Up to the RR contribution, this superpotential can be written for a constant warp factor as
\beq
W_{{\rm NS}} = C \int_{\mmm} \langle \P_1^0 , \left(\d -  H\w\right) \left(e^{-\p} \im \P_2^0\right) \rangle \ , \label{W}
\eeq
with a constant $C$ and the Mukai product defined in \eqref{Mukai}. Let us comment on this expression. For a supersymmetric Mink vacuum without RR, $\left(\d -  H\w\right) \left(e^{-\p} \im \P_2^0\right)$ vanishes; in case RR are present, their contribution to the superpotential is also precisely the one that cancels \eqref{dP2stand0}. So the formula \eqref{W} gives $W=0$ as expected. Another way to see this is to use \cite{Grana:2006kf}
\beq
\int_{\mmm} \langle \Psi_1 ,  \left(\d -  H\w\right) \Psi_2 \rangle = \int_{\mmm} \langle \left(\d -  H\w\right) \Psi_1 , \Psi_2 \rangle \ , \label{WLeib}
\eeq
to rather get $\left(\d -  H\w\right) \P_1^0$: the other condition \eqref{dP1stand0} makes again the superpotential vanish for a Mink vacuum, up to a derivative of the dilaton. This last derivative should however not contribute because of the compatibility condition \eqref{compat0}, as we will see below.\footnote{Note also that a constant warp factor typically leads to a constant dilaton in the (SUSY) vacuum.} Finally, for an AdS vacuum, one gets from \eqref{dP1stand0} or \eqref{dP2stand0} that $W$ is related to $\mu$, so to the cosmological constant, as expected \cite{Koerber:2007xk}. Let us mention as well that $W_{{\rm NS}}$ can be rewritten without $H$ but with $e^{-b}$ on both pure spinors, thanks to \eqref{Deb}, leading for an SU(3) structure to the standard $J_c=b+\i J$ combination.

\subsection{Proposed superpotential and comparison to the literature}

The formula \eqref{W} was extended in \cite{Micu:2007rd, Ihl:2007ah, Cassani:2007pq} to include non-geometric fluxes. Performing a dimensional reduction, these papers further compared their $W$ to corresponding four-dimensional superpotentials expressed in terms of moduli \cite{Shelton:2005cf, Aldazabal:2006up}. To include $Q$- and $R$-fluxes, the idea was to replace $\d -  H\w$ by a more general derivative operator, such as the $\D_{\sharp}$ of \cite{Ihl:2007ah} discussed in details in \cite{Andriot:2014uda}. So we naturally propose here for the (NSNS) superpotential (with constant warp factor)
\beq
\tilde{W}_{{\rm NS}} = \frac{C}{2} \int_{\mmm} \langle e^{-\tp}\P_1^0 , \D \im \P_2^0 \rangle \ , \label{Wgen}
\eeq
as already mentioned in \eqref{Wgenintro}; $\D$ is the Dirac operator discussed previously. Picking for the latter the standard supergravity one \eqref{Diracstandard}, the general formula \eqref{Wgen} reproduces the standard $W$ \eqref{W}. Doing the same in $\b$-supergravity leads to an expression for $W$ with non-geometric fluxes. A difference with previous papers is that $Q$ and $R$ have here a ten-dimensional interpretation. We now compute this superpotential more explicitly for an SU(3) structure, and compare our results to formulas of the literature.

We use the expression \eqref{Diracflat} for the Dirac operator in $\b$-supergravity, and the various definitions. We also consider that the coefficients in flat indices of the SU(3) structure forms, $J_{ab}$ and $\Omega_{abc}$, do not depend on internal coordinates: these coefficients are usually replaced by moduli, that only have a four-dimensional dependence. Finally, we recall that $J$ is a real two-form. In type IIB, we obtain at first
\bea
\tilde{W}_{{\rm NS}} = \i C \int_{\mmm} e^{-\tp} \Big(& c_\theta\ f \diamond J + \frac{s_\theta}{2}\ Q \diamond ( J\w J) +\frac{c_\theta}{3!}\ R \vee (J\w J \w J) \label{WIIB}\\
& + c_\theta\ \d\tp \w J + \frac{s_\theta}{2}\ (\tau - \cN \tp) \vee (J\w J) \Big) \w \Omega \nn \ ,
\eea
with $c_\theta= \cos(\theta_+)$, $s_\theta= \sin(\theta_+)$. More explicitly,
\bea
& f \diamond J = \frac{1}{2} f^{c}{}_{ab}  J_{ce}\ \te^{a}\!\w \te^{b}\!\w \te^e \\
& \frac{1}{2}\ Q \diamond ( J\w J) = \frac{1}{2} Q_{a}{}^{bc} \left( \frac{1}{2} J_{cb}J_{ef} - J_{ce}J_{bf} \right) \te^a\!\w \te^e\!\w\te^f\nn\\
& \frac{1}{3!}\ R \vee (J\w J \w J) =  \frac{1}{2} R^{abc} \left( \frac{1}{2}  J_{cb}J_{ae}J_{fg} -\frac{1}{3}  J_{ce}J_{bf}J_{ag} \right) \te^e\!\w\te^f\!\w\te^g\nn\\
& \frac{1}{2}\ (\tau - \cN \tp)  \vee (J\w J) = \frac{1}{2} \left( - \frac{1}{2}\b^{bc}f^{a}{}_{bc} + \b^{ab}\del_b \tp \right)  J_{ae}J_{fg}\ \te^e\!\w\te^f\!\w\te^g\nn \ .
\eea
The last equation indicates that $(\tau - \cN \tp)  \vee (J\w J)$ is proportional to $J$. This implies that the second row of \eqref{WIIB} is proportional to $J\w \Omega$. Requiring an SU(3) structure, i.e. enforcing the compatibility condition \eqref{compat}, then makes this second row vanish.\footnote{This reasoning could be generalized to a generic SU(3)$\times$SU(3) structure with the condition \eqref{compat0}, allowing to discard the dilaton terms, or more precisely the terms due to the $\mathbb{R}^+$ factor. The corresponding terms at the level of DFT are also truncated in \cite{Blumenhagen:2013hva} thanks to another argument: the orientifold projection, considered odd on the winding. Interestingly, we do not need this projection so far here.} Note this property needs to remain true despite the moduli fluctuations. Using further \eqref{compat}, only some terms remain from the $Q$- and $R$-fluxes contributions: we get effectively the following reduced actions
\bea
& \frac{1}{2}\ Q \diamond_r ( J\w J) = -\frac{1}{2} Q_{a}{}^{bc} J_{ce}J_{bf}  \te^a\!\w \te^e\!\w\te^f \\
& \frac{1}{3!}\ R \vee_r (J\w J \w J) = -\frac{1}{3!} R^{abc}  J_{ce}J_{bf}J_{ag} \te^e\!\w\te^f\!\w\te^g\ .
\eea
Finally, let us distinguish between the possible phases according to the choice of orientifold. Note that the notion of orientifold is not really defined in the context of $\beta$-supergravity, since the RR sector has not been studied so far; the distinction between O3, O7, and O5, O9, should then be viewed more formally as a choice on the phase $\theta_+$.\footnote{\label{foot:O9het}The O5, O9, case is also sometimes discussed, in the literature on superpotentials, with heterotic, as the two are simply related by S-duality.} Absorbing the possible minus sign of \eqref{phasechoice} in a redefinition of $C$, we get
\bea
\mbox{O3 or O7:}\quad &\tilde{W}_{{\rm NS}} = \i C \int_{\mmm} e^{-\tp} \left( \frac{1}{2}\ Q \diamond_r ( J\w J) \right) \w \Omega \ , \label{WO3} \\
\mbox{O5 or O9:}\quad &\tilde{W}_{{\rm NS}} = \i C \int_{\mmm} e^{-\tp} \left( f \diamond J + \frac{1}{3!}\ R \vee_r (J\w J \w J) \right) \w \Omega \ . \label{WO5}
\eea
Let us compare these formulas to those in the literature. The first superpotential with non-geometric fluxes was proposed in \cite{Shelton:2005cf} based on duality arguments, and was given in terms of moduli (STU model). This expression was recovered in \cite{Aldazabal:2006up} for type IIB with an O3-plane from an expression in terms of internal forms; we compare the latter to our \eqref{WO3}. Forgetting about the $H$-flux, we find an exact agreement, fixing $C=-\frac{1}{3}$. The same type of contraction as $Q \diamond$ appears in the superpotential of \cite{Aldazabal:2006up}, as well as in \cite{Blumenhagen:2013hva}. For the case of an O5- or O9-plane, we have not found in the literature an expression in terms of forms to be compared to our \eqref{WO5}; this expression is then new to the best of our knowledge. It could be used as well for heterotic (see footnote \ref{foot:O9het}). One should still perform the expansion and integration on a form basis to get the expression in terms of moduli. This is however beyond the scope of this paper. From T-duality arguments, our expression \eqref{WO5} looks in any case very plausible.

We now turn to type IIA, that should be compared in the literature to the case of an O6-plane. Our formula \eqref{Wgen} leads to a $\D \re \Omega$. This quantity however does not appear in the literature, except in \cite{Grana:2005ny} but without non-geometric fluxes, and in \cite{Cassani:2007pq} where it is only implicitly proposed. On the contrary, $\D$ is rather acting on $\Phi_+^0$ in \cite{Aldazabal:2006up} and \cite{Ihl:2007ah}. Such a situation could only be reached after integrating by parts with $\D$, similarly to \eqref{WLeib}. While the latter holds for standard supergravity thanks to the absence of boundary on the compact $\mmm$ and the $H$-flux acting with a wedge, the analogous result for $\D$ in $\b$-supergravity is not obvious to derive, because of the contractions on forms. A proof might still be obtained using the specific form \eqref{Debetaintro} of $\D$, or that $\D$, a Dirac operator, acts on pure spinors, or comparing expressions of the superpotential in terms of moduli. In any case, let us assume here that this property holds, i.e.
\beq
\int_{\mmm} \langle \Psi_1 , \D \Psi_2 \rangle = \int_{\mmm} \langle \D \Psi_1 , \Psi_2 \rangle \ ,
\eeq
allowing us to start in type IIA with
\beq
\tilde{W}_{{\rm NS}} = \frac{C}{2} \int_{\mmm} \langle \D( e^{-\tp}\P_1^0) , \im \P_2^0 \rangle \ . \label{WgenIIA}
\eeq
Pursuing the same reasoning as in type IIB, we derive from \eqref{WgenIIA}
\beq
\tilde{W}_{{\rm NS}} = -e^{\i \theta_+} C \int_{\mmm} e^{-\tp} \Big( \i\ f \diamond J + \frac{1}{2}\ Q \diamond_r ( J\w J) +\frac{\i}{3!}\ R \vee_r (J\w J \w J) \Big) \w \re \Omega \ . \label{WIIA}
\eeq
This formula agrees completely with the proposal of \cite{Ihl:2007ah}, up to fixing $C$. The same goes for the comparison to \cite{Blumenhagen:2013hva}, up to a redefinition of $\Omega$, and a conventional minus sign difference in the $R$-flux. Finally, our formula agrees with that of \cite{Aldazabal:2006up}, up to $C$ and numerical factors in the contractions.

The comparison of our proposed $W$ and formulas of the literature implicitly considers that the ten-dimensional non-geometric fluxes of $\b$-supergravity are the same as the four-dimensional ones. This has worked well so far, but in type IIA, we did not reach formulas with explicit moduli dependence.\footnote{Doing so would allow a comparison to \cite{Derendinger:2014wwa, Danielsson:2014ria} where new non-geometric terms were obtained from M-theory. They are however unlikely to be reproduced here, due to the assumption of an SU(3) structure.} Indeed, a derivation of the moduli formula of \cite{Shelton:2005cf} does not seem to have been performed directly in the literature, its comparison to other expressions is usually rather done thanks to duality arguments. In \cite{Blumenhagen:2013hva}, an oxidation is made in type IIA, instead of a reduction, and ends with a comparison and matching of the DFT Lagrangian of \cite{Geissbuhler:2013uka}. Since $\b$-supergravity fluxes (and Lagrangian) agree with the DFT ones, upon the strong constraint and setting $b=0$ \cite{Andriot:2013xca}, we would conclude on a matching with \cite{Blumenhagen:2013hva}. In the latter however is indicated a difference between ten-dimensional and four-dimensional fluxes, on the contrary to what we have considered so far. This discrepancy might be related to the way four-dimensional scalar fields, loosely called here moduli, are defined. Following STU models, the authors of \cite{Blumenhagen:2013hva} include the fluctuation of the $b$-field in a modulus; in $\b$-supergravity, we would obviously not get such a modulus when expanding of our superpotential. It is unclear whether the $b$-field modulus would simply be traded for us into a $\b$ modulus, because there is actually no explicit dependence in $\b$ in \eqref{WO3}, \eqref{WO5} and \eqref{WIIA}. An expansion of our superpotential may then only include the geometric moduli and dilaton, and the comparison should thus be done at that level. These points deserve in any case more study. Still, we conclude that the general formula(s) proposed here for the superpotential, depending on the Dirac operator, reproduces remarkably well expressions in the literature in terms of structure forms and non-geometric fluxes.

\section{Geometrical characterisation and more on the $\b$-twist}\label{sec:more}

For both standard and $\b$-supergravity, the Dirac operator \eqref{Diracstandard} or \eqref{Diraccurved} can be rewritten in terms of exponentials, either of the $b$-field \eqref{Deb} or of $\b$ \eqref{Debetaintro}. The rewriting of the former is straightforward, and we prove the latter in appendix \ref{ap:betatwist}; let us recall this result here
\beq
\D A= 2 e^{\tp} e^{\b\vee} \d (e^{-\b\vee} e^{-\tp} A)\ .\label{Debeta}
\eeq
As discussed in the Introduction, these exponentials can be viewed as twists on $T\mmm\oplus T^*\mmm$, already by looking at the generalized vielbein $\eee$ or $\teee$ \eqref{genvielb} corresponding to each theory. We discuss this point in more details in the following, and argue how this is crucially related to the geometrical characterisation of (SUSY) backgrounds of $\b$-supergravity. We further elaborate on the consequences of this $\b$-twist, for the RR sector, and in relation to intermediate and dynamical SU(2) structure solutions.

\subsection{Geometrical characterisation of the backgrounds}\label{sec:geomcharac}

Conditions for preserving SUSY usually provide a geometrical characterisation of the manifold $\mmm$, the prime example being the Calabi-Yau. As mentioned in the Introduction, the formulation in terms of GCG \cite{Hitchin:2004ut, Gualtieri:2003dx} has provided such a characterisation in presence of background fluxes (see \cite{Grana:2006kf, Koerber:2010bx} for reviews). We analyse in this section the situation for $\b$-supergravity. Let us first recall some terminology, and the results for standard supergravity. To each pure spinor $\Phi$ (here of non-zero norm) corresponds a generalized complex structure (GCS). Having a pure spinor satisfying
\beq
\d \Phi = (v \vee + \xi \w) \Phi \ , \label{GCS}
\eeq
for some $(v+\xi) \in T\mmm\oplus T^*\mmm $ is equivalent to its GCS being integrable: $\mmm$ is then generalized complex. Furthermore, if $\Phi$ is closed, $\mmm$ is a generalized Calabi-Yau (GCY). Finally, having a generalized K\"ahler manifold requires two distinct closed pure spinors. For standard supergravity, a SUSY Mink background with $H=0$ asks for $\mmm$ to be a GCY \cite{Grana:2005sn, Grana:2006kf}: the pure spinor $e^{2A-\phi} \Phi_1$ in \eqref{dP1stand} is closed for $\mu=0$ and $H=0$. In absence of RR fluxes, with a constant warp factor, the second condition \eqref{dP2stand} further constrains to a generalized K\"ahler manifold (reviews on this case can be found in \cite{Gualtieri:2010fd, Sevrin:2013oca, Terryn:2013kr}). The GCY characterisation was proven useful, leading for instance to an extensive search for solutions on six-dimensional nilmanifolds, as those are all GCY \cite{Cavalcanti}. In presence of a closed $H$-flux, the corresponding $b$-field induces a twist; a pure spinor closed under $d-H\w$, as in \eqref{dP1stand} with $\mu=0$, then characterises a twisted GCY. The twist by the $b$-field can be seen through the rewriting $d-H\w= e^{b\w} \d e^{-b\w}$, or in the off-diagonal block of the generalized vielbein $\eee$ \eqref{genvielb}, as discussed in the Introduction. It twists the local $T\mmm\oplus T^*\mmm $ into the, globally non-trivial, generalized tangent bundle $E_T$.

We now turn to $\b$-supergravity and the pure spinors conditions \eqref{dP1} and \eqref{dP2}, to study whether an analogous characterisation can be obtained.\footnote{For an SU(3) structure, an alternative might be to study the conditions in terms of the SU(3) torsion classes, and compare them to the fluxes, as e.g. in \cite{Grana:2005jc}.} We focus on the Mink case ($\mu=0$) and the first condition \eqref{dP1}; the second one is expected to be corrected by a RR contribution. We thus look at \eqref{dP1M}, written with the Dirac operator \eqref{Debeta} as
\beq
\D \P_1 = - 4 \left( \d A \w  + \cN A \vee \right) \P_1  \label{dP1M2} \ .
\eeq
This equation is analogous to the case of a ($b$-twisted) integrable GCS, as in \eqref{GCS}. As mentioned in section \ref{sec:purespin}, whenever $\cN A=0$, the warp factor can be absorbed in the LHS, to get $e^{2A} \P_1$ closed under $\D$, precisely as for standard supergravity. We now consider this case in more details, i.e.
\beq
\D \Phi =0 \ . \label{DP}
\eeq
This is the analogue to the $b$-twisted GCY condition. In the Dirac operator acting on the pure spinor, the standard $e^{b\w}$ is here traded for $e^{\b\vee}$, as given in \eqref{Debeta}, precisely as the corresponding generalized vielbein $\teee$ \eqref{genvielb} now has $\b$ in the (other) off-diagonal component. This can also be understood when viewing these exponentials as O(d,d) elements acting in the spinorial representation on the spinors, with $\G^{{\cal A}}$ given by forms and contractions, while $\eee$ and $\teee$ are correspondingly O(d,d) matrices (see e.g. \cite{Andriot:2009fp}). It is thus natural in $\b$-supergravity to talk of a twist by $\b$, and to consider \eqref{DP} as a $\b$-twisted GCY condition; for $\b=0$, we recover a GCY condition. If we can make sense of it, the geometrical characterisation of $\mmm$ is then this $\b$-twisted GCY. Furthermore, in absence of RR-fluxes and with a constant warp factor (the latter is automatic for a SUSY solution, from \eqref{susy1}), we get from \eqref{dP2} a second pure spinor closed under $\D$, analogously to a twisted generalized K\"ahler $\mmm$.

Under the $\b$-twist, the local $T\mmm\oplus T^*\mmm $ should be globally described by a generalized cotangent bundle $E_{T^*}$, discussed in \cite{Andriot:2013xca} (see also \cite{Grana:2008yw}), and given by
\beq
\begin{array}{ccc} T \mmm & \hookrightarrow & E_{T^*} \\
 & & \downarrow \\
 & & T^* \mmm
\end{array} \label{ET*}
\eeq
i.e. fibered reverse wise with respect to the standard $E_T$. Characterising precisely $\mmm$ in $\b$-supergravity as a $\b$-twisted GCY amounts to define properly this $E_{T^*}$. To do so, there should be global conditions and restrictions on the way $\b$ is patched, analogously to the cocycle conditions for the $b$-field and $E_T$, as discussed in the Introduction. In addition, for the bundle $E_{T^*}$ to be physically relevant for $\b$-supergravity, these patching transformations of $\b$ (and of the other fields) should be symmetries of the theory. This was studied in details in \cite{Andriot:2013xca, Andriot:2014uda}: we found that in general, symmetries of $\b$-supergravity are not suited to define $E_{T^*}$; however, by considering restrictions or subcases, we showed that a symmetry enhancement could occur (as assuming isometries provides T-duality to string theory), providing new symmetries allowing to construct $E_{T^*}$. We determined in \cite{Andriot:2014uda} a class of backgrounds where this scenario is realised. Those admit $n$ isometries generated by constant Killing vectors, i.e. all fields are independent of $n$ coordinates. In this subcase, the transformation
\beq
\forall m,p,q,\ \b^{pq} \rightarrow \b^{pq} + \varpi^{pq},\ \del_m \varpi^{pq} =0,\ \varpi^{pq}=0\ \mbox{for $p$ or $q$ not along the $n$ isometries} \label{betatransfo}
\eeq
with $\varpi$ antisymmetric, is a (manifest) symmetry of $\b$-supergravity, that leaves the non-geometric fluxes invariant. It is called the $\b$-transform, and is an element of the T-duality group $O(n,n)$. Backgrounds of $\b$-supergravity where fields are patched using $\b$-transforms and diffeomorphisms were shown to be globally well-defined and geometric \cite{Andriot:2014uda}; $E_{T^*}$ is then likely to be defined without obstruction. In addition, such backgrounds were shown to correspond in standard supergravity to non-geometric backgrounds, themselves T-dual to geometric ones. The latter admit by definition a standard $E_T$ description, so the $E_{T^*}$ discussed here can be viewed as an alternative and dual description of an existing $E_T$. It is traditionally considered that non-geometric backgrounds cannot be described by GCG and $E_T$ (see e.g. \cite{Belov:2007qj, Grana:2008yw, Coimbra:2011nw} for obstructions), but the construction just mentioned now seems to provide a description within Generalized Geometry, understanding this formalism in a slight extended sense though. An important remaining question is whether there are other cases than the one detailed above for which constructing $E_{T^*}$ is possible; if so, the hope is that those other backgrounds would provide truly new physics \cite{Andriot:2014uda}. We sketch in appendix \ref{ap:invD} an argument that concludes rather negatively on this question.

Let us finally go beyond the structures proposed so far to characterise the backgrounds, and mention other possibilities pointed out recently in the literature. To start with, bivectors enter in Poisson geometry: for instance, $\b$ defining a Poisson structure is equivalent to a vanishing $R$-flux.\footnote{Let us make a side remark. The commutator of two $\cN$ on a vector $V$ is given by the $R$-flux and the analogue for $\cN$ of the Riemann tensor, namely $\cR^{mn}{}_p{}^q$ \cite{Andriot:2012an, Andriot:2013xca}. Considering the square of $\iota_m \cN^m \cdot $ on an object $V^p\iota_p$ is then given by the $R$-flux and $\cR^{[mn}{}_p{}^{q]}$, where the latter is actually related to $\N_p R^{mnq}$. Therefore, for a Poisson $\b$, the operator $\iota_m \cN^m \cdot $ squares to zero on contractions, analogously to the exterior derivative $\d$ on forms. A corresponding cohomology could then be studied. Here though, this operator rather acts on forms.} This could be an interesting subcase for $\b$-supergravity. For Poisson structures in GCG, see references in \cite{Chen}. A related object is the Schouten-Nijenhuis bracket, defined with a bivector $\pi$: $[\pi, \cdot ]_{{\rm SN}}$. A differential $\d_{\pi}$ acting on a vector can also be defined, as being equal to the bracket action (see e.g. \cite{Blumenhagen:2012nt}). The quantity $[\pi, \pi]_{{\rm SN}}$ is given by the $R$-flux, with $\pi$ instead $\b$: its vanishing is thus equivalent to $\pi$ defining a Poisson structure. It is also equivalent to $(\d_{\pi})^2=0$, from the Jacobi identity. For $\b$-supergravity, one may choose $\pi=\b$, or stay more general and consider the two bivectors separately. For instance, if $\mmm$ admits a symplectic structure, it admits most of the time as well a Poisson structure (taking, when possible, the inverse of the symplectic form to be the Poisson bivector). In that case, one can have two bivectors: the Poisson $\pi$ and the dynamical field $\b$. This could provide more structures: for example, one could study whether the transformation $\b \rightarrow \b + \d_{\pi} V$, with a vector $V$, is a symmetry of $\b$-supergravity. Further structures have been indicated in the literature. Exponentials $e^{\theta \vee}$ for a bivector $\theta$, similar to the exponential in \eqref{Debeta}, enter in several mathematical contexts, in particular as $\b$-transforms acting on forms. Dirac structures and related algebroids, or deformations of them, have been defined from such exponentials as well as $b$-transforms \cite{Chatzistavrakidis:2013wra, Asakawa:2014eva, Brahic:2014bta}.\footnote{In \cite{Asakawa:2014eva}, these structures are related to $D$-branes. Combining that work and our results could then lead to a calibration of SUSY branes with pure spinors, in $\b$-supergravity.} Lie and Courant algebroid structures involving a bivector were also introduced and studied in \cite{Blumenhagen:2012pc, Blumenhagen:2012nk, Blumenhagen:2012nt, Blumenhagen:2013aia}. There, a symmetry named $\b$-diffeomorphism was considered. In a different though related construction \cite{Asakawa:2014kua}, a new bundle similar to our generalized cotangent bundle $E_{T^*}$ was proposed; it defines a "Poisson-generalized geometry", where the patching is done via $\b$-diffeomorphisms. Unfortunately, neither of these constructions seems to provide a geometrical description of $\b$-supergravity. Indeed, $\b$-diffeomorphisms are not a symmetry of the latter, if applied literally on $\b$: a reason for this is that the field redefinition giving $\b$ from the standard supergravity $g$ and $b$ is in \cite{Blumenhagen:2012nk, Blumenhagen:2012nt} not the same as \eqref{fieldredef}. Furthermore, two distinct bivectors are considered in \cite{Asakawa:2014kua}, one with vanishing $R$-flux: this does not apply to $\b$-supergravity in full generality, but could be considered in a particular case, as discussed previously. One may still wonder whether $E_{T^*}$ could be defined as an algebroid, and what bracket should be associated to it. It could be worth studying at first whether the Courant bracket (or a $\b$-twist version of it) is preserved under $\b$-transforms, in presence of isometries.

\subsection{On Ramond-Ramond fluxes in $\b$-supergravity}\label{sec:RR}

RR fluxes $F_p$ of standard type II supergravities often appear through the polyform $F=\sum F_p$. Forgetting about the Romans mass for simplicity, one has $F=(\d-H\w ) C$, where $C$ is again a polyform, given by a sum of corresponding potentials. Both $F$ and $C$ are known to behave as O(d,d) spinors, see \cite{Andriot:2014uda} for a discussion and references on this point. This is consistent with $\d-H\w$ being a Dirac operator (the dilaton can be considered here constant): the spinor $C$ acted on with the Dirac operator gives another spinor $F$. Related considerations appeared recently in \cite{Hohm:2011dv, Jeon:2012kd, Geissbuhler:2013uka}; see in particular \cite{Hohm:2011dv} for the spinorial representation. In $\b$-supergravity, we have a different Dirac operator. Several candidates can then be thought of for the, so far not obtained, RR fluxes:
\begin{itemize}
\item A first possibility is simply $e^{\b\vee} \d (e^{-\b\vee} C)$. This would essentially result in non-geometric NSNS fluxes contracting on standard RR potentials. These contractions may either lead to new types of RR fluxes, or rather complete fluxes of standard supergravity. In a supergravity theory containing both a $b$-field and a $\b$, with a constraint among them, as discussed e.g. in \cite{Andriot:2013xca}, the Dirac operator would a priori depend on all NSNS fluxes, and applying it on $C$ would lead to a completion of the standard RR fluxes, as in \cite{Blumenhagen:2013hva}. The above possibility $e^{\b\vee} \d (e^{-\b\vee} C)$ would then correspond to a subcase in this more general setting.

\item Another possibility is to have a different kind of potentials, given by polyvectors instead of forms \cite{Andriot:2013xca}. A sum of those denoted by $\gamma$ could correspond to another O(d,d) spinor. The (sum of) RR fluxes would then be given by $e^{\b\vee} \d (e^{-\b\vee} \gamma)$, which would certainly provide new types of fluxes. This proposal was sketched in \cite{Aldazabal:2010ef} where the first terms of the expressions for these new fluxes were given from deformations. In \cite{Andriot:2010sya}, $\b$ was introduced by picking the generalized vielbein $\teee$ \eqref{genvielb} instead of the standard $\eee$; the RR polyvector potentials just mentioned could appear similarly in a setting where the RR sector is captured by the generalized vielbein, i.e. exceptional generalized geometry or field theory. Choosing there a different generalized vielbein, or equivalently a different group generator in the representations decomposition, instead of the standard one of e.g. \cite{Hohm:2013uia, Aldazabal:2013via}, could provide a derivation of the RR fluxes (see also \cite{Malek:2012pw, Blair:2013gqa}). The latter could then be compared to the above expression.
\end{itemize}
RR fluxes of $\b$-supergravity may provide an uplift to some of the known four-dimensional RR non-geometric fluxes \cite{Aldazabal:2006up, Aldazabal:2008zza, Dibitetto:2010rg}. The former would then be new types of fluxes, which is in agreement with the expressions proposed above. Reproducing four-dimensional fluxes is however not the decisive criterion. Rather, the RR fluxes should ensure consistency of the $\b$-supergravity theory, as in standard supergravity. One way to make them appear consistently would be through a field redefinition from the standard RR fields, similarly to the NSNS sector.\footnote{In the world-sheet approach by Berkovits, named the pure spinor formalism, the standard RR fluxes enter explicitly, and appear again as an O(d,d) spinor. Changing that spinor, or that of the RR potentials, is one possibility mentioned above, and could correspond to a field redefinition. Note that the NSNS field redefinition \eqref{fieldredef} is also straightforward on the world-sheet, since it involves precisely the combination $g+b$.} It would be interesting to get the corresponding rewriting of the action, and in particular of the standard topological terms involving the $b$-field. Another way is through supersymmetry; in particular RR fluxes are expected to contribute to the RHS of \eqref{dP2}. If they are given in terms of polyvectors through the above $\gamma$, it is yet unclear how to obtain a polyform expression for them that would fit on the RHS of \eqref{dP2}. The correct formulation of the pure spinors conditions may then not be in terms of forms, but rather with O(d,d) spinors and Cliff(d,d) $\G$-matrices.

\subsection{$\b$-twist and intermediate or dynamical SU(2) structure solutions}\label{sec:dynSU2}

When looking for solutions to the pure spinors conditions, it is convenient to distinguish the three cases of an SU(3), orthogonal SU(2) or intermediate SU(2) structure, as discussed in section \ref{sec:purespinWstand}, each of them being described by a different pair of pure spinors. Those depend on parameters, namely the "moduli" (radii, complex structure moduli), the phase $\theta_+$, and for an intermediate SU(2) structure the angle between the internal spinors $\eta_+^{1,2}$. There are in addition the norms of the spinors, which are related in presence of an orientifold to the warp factor. For simplicity, the previous parameters are often set to constants when looking for solutions. It is however more interesting to get solutions with varying parameters. A first example are Frey-Gra\~na solutions \cite{Frey:2003sd}: those admit an SU(3) structure and allow for the phase $\theta_+$ to vary. More challenging are dynamical SU(2) structure solutions, for which the angle between the internal spinors varies. They can lead to genuinely SU(3)$\times$SU(3) structure solutions, that interpolate between different structures, for instance having an intermediate SU(2) structure at most points which becomes an SU(3) structure at some loci where the angle between the spinors vanishes.\footnote{Such a situation can lead to type changing loci, discussed in \cite{Torres1, Torres2, Sevrin:2011mc, Terryn:2013kr} and references therein; for type change with $\b$-transformations, see also \cite{Cavalcanti}.} SUSY Mink solutions with intermediate SU(2) structure on a compact $\mmm$ have been found in \cite{Koerber:2007hd, Andriot:2008va} (see also \cite{Andriot:2009rc, Andriot:2010sya} for clearer formulations). No SUSY solution with such a structure is allowed on AdS${}_4$ \cite{Caviezel:2008ik, Solard:2013fva}. The differential conditions coming from ${\cal N}=1$ SUSY require $\mmm$ to have some geometric properties, and those differ for the two SU(2) structure cases; the geometry underlying intermediate SU(2) structure solutions has been characterised in \cite{Fino:2010dx}. For a Mink dynamical SU(2) structure, the SUSY conditions were given in terms of forms in \cite{Andriot:2010sya}. It is notoriously difficult to get such a solution on a compact $\mmm$ and none have been found so far. One reason is the coordinate dependence together with the compactness global constraint. Dynamical SU(2) structure solutions have been found as a non-geometric background \cite{McOrist:2010jw} (see the related footnote \ref{foot:SU2dyn}), or on non-compact spaces \cite{Minasian:2006hv, Butti:2007aq, Heidenreich:2010ad, Caceres:2014uoa}; they have been further studied in \cite{Heidenreich:2011ez, Gaillard:2013vsa}.\footnote{Analogous considerations were made in M-theory compactified to three dimensions in \cite{Condeescu:2013hya, Babalic:2014cea}; relations to $D7$ gaugino-condensation are also mentioned there. In addition, a dynamical SU(3) structure solution on a seven-dimensional manifold in type IIA is mentioned in \cite{Macpherson:2013zba}.}

In section 3.5 of \cite{Andriot:2010sya}, solutions with an intermediate SU(2) structure, denoted $(\angle)$, were related by a $\b$-transform to solutions with an SU(3) structure, denoted $(||)$. As discussed there, provided a set of conditions on the fields, in particular $b_{(||)}=0$, one would have
\beq
e^{-\b\vee}\ e^{-\p_{(||)}} \Phi_{1,2 {(||)}} = e^{-b_{(\angle)}\! \w} e^{-\p_{(\angle)}} \Phi_{1,2 (\angle)}  \ . \label{relSU3SU2}
\eeq
This was shown to hold for pairs of pure spinors on a torus, as well as for those of solutions found in \cite{Grana:2006kf, Andriot:2008va} on a nilmanifold. In both cases, the angle between the internal spinors of the $(\angle)$ solution was, on the LHS of \eqref{relSU3SU2}, encoded in $\b$. Suppose now that the angle between the two spinors varies on the manifold. The (local) intermediate SU(2) structure is by definition rather a dynamical SU(2) structure. If the angle is still encoded in a $\b$ through the relation \eqref{relSU3SU2}, that $\b$ is also varying.\footnote{The relation \eqref{relSU3SU2} was shown to hold in \cite{Butti:2007aq} where the RHS was a dynamical SU(2) structure solution. The $\b$ was however constant and the dynamic was encoded in another function. More precisely, the RHS was the Lunin-Maldacena (LM) background \cite{Lunin:2005jy}. It is given by a (constant) $\b$-transform, equivalent to a combination of two T-dualities and a rotation, applied to the sphere of the standard AdS$_5 \times$S$^5$ background \cite{Minasian:2006hv, Butti:2007aq, Grana:2008yw}. LM is the gravity dual to the so-called $\b$-deformation of ${\cal N}=4$ SYM, a marginal deformation of the theory that reduces the number of SUSY.} The latter could then be interpreted as the field of $\b$-supergravity, providing a new understanding of a dynamical SU(2) structure.\footnote{The non-abelian T-duality transformation described in \cite{Barranco:2013fza} looks analogous to \eqref{relSU3SU2}; it would be interesting to compare the two, given our discussion and the result of \cite{Caceres:2014uoa}.} Let us reformulate this idea more precisely.

We recall that the Dirac operator of both standard and $\b$-supergravity can be written in terms of twists by $b$ or $\b$, as in \eqref{Deb} and \eqref{Debeta}. Consider now the exterior derivative $\d$ applied on both sides of the relation \eqref{relSU3SU2}, and set it equal to zero: the LHS corresponds to a closed pure spinor (or say, solution) of $\b$-supergravity, due to the $\b$ and \eqref{Debeta}, while the RHS is a solution of standard supergravity, due to the $b$-field and \eqref{Deb}. More precisely, suppose we have an SU(3) structure solution to the pure spinors conditions of $\b$-supergravity on a manifold $\mmm$. Assume at first its $\b$ to be constant. Then, if a relation \eqref{relSU3SU2} is established, it implies that $\mmm$ can admit as well an intermediate SU(2) structure SUSY solution of {\it standard} supergravity, because of $b_{(\angle)}$ a priori non-zero. More interestingly, if $\b$ varies (more probable for non-zero non-geometric fluxes), one gets on the RHS a dynamical SU(2) structure solution of standard supergravity. This would be an interesting way to generate and interpret such solutions. However, global aspects should be further studied. Compactness is not guaranteed in this process, global requirements on $\b$ in $\b$-supergravity might differ from those on the angle in the dynamical SU(2) structure. Constraints due to orientifolds in standard supergravity should also be taken care of. It would still be interesting to study whether the dynamical SU(2) structure solutions mentioned above, namely \cite{McOrist:2010jw} and \cite{Minasian:2006hv, Butti:2007aq, Heidenreich:2010ad, Caceres:2014uoa}, satisfy \eqref{relSU3SU2} and verify this scenario. The reformulation of the last solutions in terms of $\b$-supergravity could then provide new holographic interpretations, since they were studied in the AdS/CFT context.

\section{Outlook}\label{sec:outlook}

Motivations for the work done in this paper and the results obtained have been presented in the Introduction. Various ideas and directions remain to be investigated: some have been discussed throughout the paper, in particular in section \ref{sec:more}, and we mention here a few more. As described in sections \ref{sec:fermsusy} and \ref{sec:compactsusy}, a first result was to deduce, from the Generalized Geometry formalism and DFT, expressions for the fermionic SUSY variations in $\b$-supergravity. It would be interesting to recover them from a (fermionic) SUSY completion of the bosonic NSNS Lagrangian at hand. A related question is whether the fermions of $\b$-supergravity are obtained via a field redefinition from those of standard supergravity, or rather correspond to different states in the spinorial representations. A similar question can be asked for the Killing spinor equations. In \cite{Marchesano:2007vw}, an expression for the $Q$-flux, for us part of $\b$-supergravity, is proposed in terms of spinors of standard supergravity in a non-geometric background. A relation between fermions of both theories would help clarifying this proposal.

An important object in this work is the Dirac operator $\D$, that acts naturally on O(d,d) spinors. It appears in the pure spinors conditions, but also in the superpotential, in the RR fluxes, in the NSNS fluxes Bianchi identities (BI), and plays an important role in the geometrical characterisation of the backgrounds. It would be interesting to determine this operator in other theories, such as the heterotic string. For the latter, pure spinors conditions were derived in \cite{Andriot:2009fp}, in an ${\cal N}=1$ Mink vacuum. An object, that could correspond to $\D$, was obtained. It contains dilaton derivatives, and the $H$-flux with a few contractions instead of wedges, similarly to the results of \cite{Grana:2004bg}, and reminiscent of what we obtained here. Note that the equation involving the $H$-flux can also be written in terms of $*H$, or with the operator $\d^c=\i (\ov{\del} - \del)$. It would then be natural that the "S-dual" equation, i.e. the condition \eqref{dP2stand} where RR fluxes enter with a Hodge star, could be written as well with $\d^c$. This rewriting was obtained in \cite{Tomasiello:2007zq}: it involves a generalization of $\d^c$ depending on a generalized complex structure ${\cal J}_{{\cal A}}{}^{{\cal B}}$. Reformulating this generalized operator in a Spin(d,d) language would be interesting (it could be given by $\G^{{\cal A}} {\cal J}_{{\cal A}}{}^{{\cal B}} D_{{\cal B}}$). This would provide to the second pure spinors condition a fully spinorial interpretation, while the first condition was already considered as a Dirac equation. This spinorial perspective could be helpful, for instance in dimensional reductions.

In standard supergravity, the pure spinors conditions have been used to various ends. For ${\cal N}=1$ Mink vacua without NS5-brane, these conditions, equivalent to preserving SUSY, were shown to imply, together with fluxes BI, that all equations of motion are satisfied \cite{Lust:2004ig, Gauntlett:2005ww, Grana:2006kf, Koerber:2007hd}. This result is an important technical simplification when looking for vacua. It would be interesting to derive the analogous result in $\b$-supergravity. This is expected to hold using Killing spinor equations instead of pure spinors conditions. Indeed, as for standard supergravity \cite{Coimbra:2011nw, Hohm:2011nu}, the $\b$-supergravity equations of motion were written in terms of the Spin(9,1)$\times$Spin(1,9) derivatives acting on a spinor \cite{Andriot:2014uda}, using the BI. If these derivatives vanish, the equations of motion are satisfied; furthermore, this vanishing is precisely the Killing spinor equations, as explained in section \ref{sec:fermsusy}, hence the result. Reaching the same conclusion from the pure spinors conditions would be interesting.\footnote{A notion of generalized special holonomy, defined from extensions of the Spin(9,1)$\times$Spin(1,9) derivatives to EGG, is considered in \cite{Coimbra:2014uxa} to characterise preserved SUSY. Expressing this notion directly in terms of our Dirac operator and Spin(d,d)$\times \mathbb{R}^+$ derivative acting on the pure spinors would be interesting as well.}

These conditions were also used for the calibration of SUSY D-branes, and similar results were obtained for NS5-branes (see \cite{Koerber:2007hd} and references therein).\footnote{Strictly speaking though, NS5-branes cannot be described in GCG since the latter requires $\d H=0$. A related discussion can be found in \cite{Coimbra:2014qaa}.} Our results may then be of interest for $5_2^2$- or Q-branes, and further exotic branes. Mutually BPS intersecting NS-branes have been considered in \cite{Hassler:2013wsa}: those could be examples to test these ideas. Our work may help as well for the world-volume actions of NS-branes, discussed e.g. in \cite{Bergshoeff:1997gy, Eyras:1998hn, Chatzistavrakidis:2013jqa, Kimura:2014upa}.

\vspace{0.4in}

\subsection*{Acknowledgments}

We would like to thank G. Aldazabal, A. Deser, D. L\"ust, R. Minasian, M. Petrini, H. Samtleben, and D. Tsimpis, for helpful discussions related to sections \ref{sec:more} and \ref{sec:outlook}. The work of D. A. is part of the Einstein Research Project "Gravitation and High Energy Physics", which is funded by the Einstein Foundation Berlin.

\newpage

\begin{appendix}

\section{Conventions}\label{ap:conv}

In this paper, the ten-dimensional flat (tangent space) indices are denoted $A \dots L$ and the curved ones are $M \dots Z$. The notation is analogous when compactifying, but we use Latin indices for the internal directions ($a\dots$ for flat, $m\dots$ for curved), and Greek indices for four-dimensional ones ($\alpha\dots$ for flat, $\mu\dots$ for curved). Finally, the index ${\cal A}$ runs over a $2d$ range for a $d$-dimensional space-time, and denotes the flat O(d,d) index. The formulas given below are written with six-dimensional indices but are actually valid in any dimension.

The vielbein $\te^a{}_m$ and its inverse $\te^n{}_b$, associated to the metric $\tg$ by $\tg_{mn}=\te^a{}_m\, \eta_{ab}\, \te^b{}_n$, allow to go from curved to flat indices. A $p$-form $A$ is given by
\beq
A=\frac{1}{p!}A_{m_1 \dots m_p}  \d x^{m_1} \w \dots \w \d x^{m_p} = \frac{1}{p!}A_{a_1 \dots a_p}  \te^{a_1} \w \dots \w \te^{a_p} \ .
\eeq
The contraction of a vector $V=V^m \del_m=V^a \del_a$ on $A$ is defined by
\beq
V \vee A= \frac{1}{(p-1)!}V^{m_1} A_{m_1 \dots m_p} \d x^{m_2} \w \dots \w \d x^{m_p} \ . \label{Contraction}
\eeq
It is also denoted by $\iota_a=\te^m{}_a \iota_m$, that satisfies the following commutation relations
\beq
V\vee A= V^a \iota_a A \ ,\qquad \{\te^{a},\iota_{b}\}=\delta^{a}_{b}\ , \quad \{\iota_{a},\iota_{b}\}=0 \ ,
\eeq
while a contraction on a scalar vanishes.

The spin connection coefficient $\o_b{}^a{}_{c}$ and the structure constant $f^{a}{}_{bc}$ (or so-called geometric flux) are defined as
\bea
& \N_b (\del_c) \equiv \o_b{}^a{}_{c} \del_a \ , \ \o_b{}^a{}_{c} \equiv \te^n{}_b \te^a{}_m \left(\del_n \te^m{}_c + \te^p{}_c \G^m_{np} \right)  = \te^n{}_b \te^a{}_m \N_n \te^m{}_c \ , \label{defspincon}\\
& f^{a}{}_{bc} = 2 \te^a{}_m \del_{[b} \te^m{}_{c]} = - 2 \te^m{}_{[c} \del_{b]} \te^a{}_{m} \label{fabc} \ ,
\eea
and the covariant derivative in flat indices on a vector $V$ is given by
\beq
\N_a V^b = \del_a V^b + \o_a{}^b{}_{c} V^c \ .
\eeq
For the Levi-Civita connection, one has the relation \eqref{of} between $\o$ and $f$. We also use the other covariant derivative $\cN$ whose action on a vector $V$ in flat indices is given by
\beq
\cN^a V^b = -\b^{ac}\del_c V^b + \co^a{}_c{}^{b} V^c \ . \label{cNvector}
\eeq
Its spin connection $\co^a{}_c{}^{b}$ was denoted ${\o_Q}_c^{ab}$ in \cite{Andriot:2013xca, Andriot:2014uda}; it has a definition analogous to \eqref{defspincon}. It is related to the $Q$-flux in \eqref{coQ}, which is similar to \eqref{of}. We refer to those two papers for more on this covariant derivative $\cN$. Its action on forms was shown to be given by
\beq
\cN^a \cdot \iota_a = -\b^{ab}\del_b\cdot\iota_{a} +\frac{1}{2} Q_{a}{}^{bc}\,\te^a\!\w \iota_b\,\iota_c -Q_{d}{}^{dc}\,\iota_c \ ,\label{cNform}
\eeq
where the dot denotes the action only on the form coefficient in flat indices.

Finally, we consider constant matrices $\gamma^a$ in flat indices, satisfying the Clifford algebra and following related properties
\bea
& \{ \gamma^a , \gamma^b \}= 2 \eta^{ab} \ , \ [ \gamma^a , \gamma^b ]= 2 \gamma^{ab} \ ,\ {\rm where} \ \gamma^{a_1 a_2 \dots a_p} \equiv \gamma^{[a_1} \gamma^{a_2} \dots \gamma^{a_p]} \  , \label{g1}\\
& \gamma^a \gamma^b = \eta^{ab} + \gamma^{ab} \ , \ \gamma^a \gamma^{bc} = \gamma^{abc} + 2 \eta^{a[b} \gamma^{c]}\ . \label{g2}
\eea

\section{Compactification and pure spinors conditions}

\subsection{Consequences of the compactification ansatz}\label{ap:compact}

From the compactification ansatz of the ten-dimensional fields given in section \ref{sec:compactsusy}, we compute here the various components of the fluxes; we recall they are defined as
\beq
f^{A}{}_{BC}=2 \te^{A}{}_{M} \del_{[B} \te^{M}{}_{C]} \ ,\ Q_{A}{}^{BC}= \del_{A} \b^{BC} - 2\b^{D[B}f^{C]}{}_{AD}\ , \ R^{ABC}= 3\te^A{}_M \te^B{}_N \te^C{}_P \b^{Q[M} \del_Q \b^{NP]} \ .\nn
\eeq
We get
\bea
& f^{\al}{}_{\b \g}=2 \te^{\al}{}_{\mu} \del_{[\b} \te^{\mu}{}_{\g]} \ ,\ f^{a}{}_{bc}=2 \te^{a}{}_{m} \del_{[b} \te^{m}{}_{c]} \ ,\ f^{\al}{}_{b \g}=- \delta^{\al}_{\g} \del_{b} A \ ,\ f^{\al}{}_{\b c}= \delta^{\al}_{\b} \del_{c} A  \ ,\\
& f^{a}{}_{\b\g}=f^{a}{}_{\b c}=f^{a}{}_{b \g} =f^{\al}{}_{b c}=0\ ,\nn\\
& Q_{a}{}^{bc}=\del_{a} \b^{bc} - 2\b^{d[b}f^{c]}{}_{ad} \ ,\ Q_{\al}{}^{b \g}= -\delta_{\al}^{\g} \b^{db} \del_{d} A \ ,\ Q_{\al}{}^{\b c}= \delta_{\al}^{\b} \b^{dc} \del_{d} A \ , \\
& Q_{\al}{}^{\b\g} =Q_{a}{}^{\b \g}=Q_{a}{}^{\b c}=Q_{a}{}^{b \g}= Q_{\al}{}^{b c}=0\ ,\nn\\
& R^{abc}= 3 \b^{d[a} \N_d \b^{bc]} \ ,\ \mbox{any other component of} \ R^{ABC}=0 \ .
\eea
From those we obtain the components of the two ten-dimensional spin connections (in flat indices)
\bea
\o_{A}{}^{B}{}_{C}\ \eta_{BD} &= \frac{1}{2}  (\eta_{BD}f^{B}{}_{AC} + \eta_{CE} f^{E}{}_{DA} + \eta_{AE} f^{E}{}_{DC})\ ,\label{of}\\
\o_{\al}{}^{\b}{}_{\g}\ \eta_{\b \delta} &= \frac{1}{2}  (\eta_{\b\delta}f^{\b}{}_{\al\g} + \eta_{\g\eps} f^{\eps}{}_{\delta\al} + \eta_{\al\eps} f^{\eps}{}_{\delta\g})\ ,\\
\o_{a}{}^{b}{}_{c}\ \eta_{bd}&= \frac{1}{2}  (\eta_{bd} f^{b}{}_{ac} + \eta_{ce}f^{e}{}_{da} + \eta_{ae} f^{e}{}_{dc}) \ ,\nn\\
\o_{\al}{}^{b}{}_{\g}\ \eta_{bd} &= - \eta_{\al \g} \del_{d} A  \ ,\ \o_{\al}{}^{\b}{}_{c}\ \eta_{\b\delta} = \eta_{\al\delta} \del_{c} A  \ ,\nn\\
\o_{\al}{}^{b}{}_{c} &= \o_{a}{}^{\b}{}_{ \g} = \o_{a}{}^{b}{}_{ \g} = \o_{a}{}^{\b}{}_{ c} =0 \ ,\nn
\eea
\bea
\co^{A}{}_{B}{}^{D}\ \eta_{DC} &= \frac{1}{2} (\eta_{CD}Q_{B}{}^{AD} + \eta_{BD} Q_{C}{}^{DA} + \eta_{BD}\eta_{CE}\eta^{AF} Q_{F}{}^{DE}) \ ,\label{coQ}\\
\co^{a}{}_{b}{}^{d}\ \eta_{dc} &= \frac{1}{2} (\eta_{cd} Q_{b}{}^{ad} + \eta_{bd} Q_{c}{}^{da} + \eta_{bd}\eta_{ce}\eta^{af} Q_{f}{}^{de}) \ ,\\
\co^{\al}{}_{b}{}^{ \delta}\ \eta_{\delta \g} &= - \delta^{\al}_{\g} \eta_{b d} \b^{ed} \del_{e} A \ , \ \co^{\al}{}_{\b}{}^{d}\ \eta_{dc} = \delta_{\b}^{\al} \eta_{cd} \b^{ed} \del_{e} A  \ ,\nn\\
\co^{\al}{}_{\b}{}^{\g} &=\co^{\al}{}_{b}{}^{c} = \co^{a}{}_{b}{}^{ \g} = \co^{a}{}_{\b}{}^{ c} =\co^{a}{}_{\b}{}^{ \g} =0 \ .\nn
\eea

As discussed in section \ref{sec:compactsusy}, the ten-dimensional $\G$-matrices satisfying the Clifford algebra $\{\G^A,\G^B\} = 2 \eta^{AB}$ are decomposed as follows
\bea
\G^A=\begin{cases}\G^{\al}=\g^{\al} \otimes \id,\quad\quad\, \al=0,\dots,3 \\ \G^a=\g_{(4)} \otimes \g^a,\quad\  a=4,\dots,9 \end{cases}\ .
\eea
The six-dimensional $\g^a$ and four-dimensional $\g^{\al}$ satisfy as well the Clifford algebra, and are constant. In addition, the $\g^a$ are purely imaginary and Hermitian: ${\g^a}^\dag = \g^a$. The chirality operators are given by $\g_{(6)}=-\i \g^{4\dots9}$ and $\g_{(4)}=\i \g^{0\dots3}$; they square to the identity and anticommute with the other $\g$-matrices. $\g_{(6)}$ is also Hermitian.

Given this decomposition, we compute the following combinations
\bea
\G^{\al \b} =& \g^{\al \b} \otimes \id \ ,\ \G^{ab} = \id \otimes \g^{ab} \ ,\\
\G^{a \b} =&\frac{1}{2}\left((\g_{(4)} \otimes \g^a)( \g^{\b} \otimes \id)-(\g^{\b} \otimes \id)(\g_{(4)} \otimes \g^a)\right)= -\g^\b \g_{(4)}\otimes \g^a \ .\nn
\eea
Together with the above components of spin connections, this leads to the following components of the spinorial covariant derivatives
\bea
&\N_{A=a}=\id \otimes \N_a \ ,\ \cN^{A=a}=\id \otimes \cN^a \ , \\
&\N_{A=\al}= \N_{\al} \otimes \id  +  \frac{1}{2}\o_{\al}{}^{b}{}_{ \g} \eta_{bd}\G^{d \g} = \N_{\al} \otimes \id  +  \frac{1}{2} \eta_{\al \b} \g^\b \g_{(4)}\otimes \g^d \del_d A \ ,\\
&\cN^{A=\al}=  \frac{1}{2} \co^{\al}{}_{c}{}^{ \b} \eta_{\b \delta} \G^{\delta c} = \frac{1}{2} \g^{\al} \g_{(4)}\otimes \g^c \eta_{c d} \b^{de} \del_{e} A \ .
\eea
These are used in the SUSY variations in section \ref{sec:compactsusy}.

\subsection{Reformulation of the supersymmetry conditions with pure spinors}\label{ap:purespin}

We introduce in section \ref{sec:purespin} the pure spinors $\P_{\pm}$ in terms of which we want to reformulate the SUSY conditions \eqref{susy1} - \eqref{susy3}. To do so, we will use the Clifford map \eqref{cliffmap}, and some of its properties that we first detail here, before starting the reformulation. For a $k$-form $A_k$, the Clifford map gives the following rules
\beq
\g^a \slashed{A}_k=\!\!\!\!\!\begin{picture}(10,10)(-10,5)
\put(0,0){\line(6,1){80}}
\end{picture}{ (\te^a \w + \eta^{ab} \iota_b) A_k}\ ,\quad  \slashed{A}_k \g^a=(-1)^{k}\!\!\!\!\!\begin{picture}(10,10)(-10,5)
\put(0,0){\line(6,1){80}}
\end{picture}{ (\te^a \w - \eta^{ab} \iota_b) A_k}\ ,
\eeq
that come from identities on $\g$-matrices, such as
\beq
\{\g^b, \g^{a_1 \dots a_k}\}=2 \g^{b a_1 \dots a_k}\ \mbox{for}\ k\  \text{even}\ ,\quad [\g^b, \g^{a_1 \dots a_k}]=2 \g^{b a_1 \dots a_k}\ \mbox{for}\ k\  \text{odd}\ .
\eeq
One subtlety in the Clifford map is due to the fact that the $\te^a$ are real while the $\g^a$ are purely imaginary. This makes a difference when considering a complex conjugation on forms of odd degree: one has
\beq
\eta^1_- \otimes \eta_-^{2 \dagger}= \overline{\sla{\P_+}}=\!\!\!\!\!\begin{picture}(10,10)(-10,5)
\put(0,0){\line(1,1){17}}\end{picture}{\overline{\P_+}}\ ,\quad \eta^1_- \otimes \eta_+^{2 \dagger}= \overline{\sla{\P_-}}=-\ \!\!\!\!\!\begin{picture}(10,10)(-10,5)
\put(0,0){\line(1,1){17}}\end{picture}{\overline{\P_-}}\ , \label{signslash}
\eeq
implying
\beq
\re( \mu\ \overline{\sla \Phi_-})= \i \hspace{-.2cm}\begin{picture}(10,10)(-10,5)
\put(0,2){\line(4,1){45}} \end{picture} \im(\overline{\mu} \Phi_-) \ .
\eeq
We now use these properties, as well as the hermitian conjugation of $\g$-matrices
\beq
\g^{a \dag} = \g^a\ ,\ \left(\g^{ab} \right)^\dag =-\g^{ab}\ ,\ \left(\g^{abc} \right)^\dag = -\g^{abc}\ ,
\eeq
to compute the exterior derivative on the pure spinors \eqref{diffspinor}. We also use the bispinor expressions, and the SUSY conditions \eqref{susy2} and \eqref{susy3} in type IIB. We obtain
\bea
2\!\!\!\!\! \begin{picture}(10,0)(-10,2)\put(0,0){\line(2,1){20}}\end{picture}{\d\P_+} =& \{ \g^a, \N_a \slashed{\P}_+\} \label{dP+Bcom}\\
=& \slashed{\N} \eta^{1}_+ \otimes \eta^{2 \dag}_+ + \g^a \eta^{1}_+ \otimes \left(\N_a\eta^{2}_+\right)^{\dagger}  + \N_a \eta^{1}_+ \otimes \eta^{2 \dag}_+ \g^a +\eta^{1}_+ \otimes \left(\slashed{\N}\eta^{2}_+\right)^{\dagger} \nn\\
=& \left(\left(\slashed{\cN}  - \frac{1}{4} \slashed{R} -\slashed{\del}\left(2A-\tp\right)+ \slashed{\b}_{\del}\left(2A+\tp\right)-\slashed{\T}\right) \eta^{1}_+ -2e^{-A}\mu\eta^{1}_-\right)\otimes \eta^{2 \dag}_+\nn\\
+& \g^a \eta^{1}_+ \otimes \left(\eta_{ab}\cN^b \eta^{2}_+ +\frac{1}{8}\eta_{ad} \eta_{be} \eta_{cf} R^{def} \g^{bc}\eta^{2}_+\right)^{\dagger} \nn\\
+& \left(- \eta_{ab}\cN^b \eta^{1}_+ +\frac{1}{8}\eta_{ad} \eta_{be} \eta_{cf} R^{def} \g^{bc}\eta^{1}_+\right) \otimes \eta^{2 \dag}_+ \g^a\nn\\
+&\eta^{1}_+ \otimes \left(\left(- \slashed{\cN}  - \frac{1}{4} \slashed{R}-\slashed{\del}\left(2A-\tp\right)- \slashed{\b}_{\del}\left(2A+\tp\right) +\slashed{\T}\right) \eta^{2}_+-2e^{-A}\mu\eta^{2}_-\right)^{\dagger}\nn\\
=& \slashed{\cN} \eta^{1}_+\otimes \eta^{2 \dag}_+ +\eta_{ab}\g^a \eta^{1}_+ \otimes \left(\cN^b \eta^{ 2 }_+ \right)^{\dag} - \eta_{ab}\cN^b \eta^{1}_+\otimes \eta^{2 \dag}_+ \g^a -\eta^{1}_+ \otimes \left(\slashed{\cN}\eta^{2}_+ \right)^{\dag}\nn\\
-& \frac{1}{4} \slashed{R}  \eta^{1}_+\otimes \eta^{2 \dag}_+ + \frac{1}{4} \eta^{1}_+ \otimes \eta^{2 \dag}_+\slashed{R} -\frac{1}{8} \eta_{ad}\g^a \eta^{1}_+ \otimes  \eta^{2 \dag}_+  \g^{bc} \eta_{be} \eta_{cf} R^{def} +\frac{1}{8} \eta_{ad} \eta_{be} \eta_{cf} R^{def} \g^{bc}\eta^{1}_+\otimes \eta^{2 \dag}_+ \g^a\nn\\
-&\lbrace \slashed{\del}\left(2A-\tp\right),\slashed{\Phi}_+\rbrace + \lbrack \slashed{\b}_{\del}\left(2A+\tp\right)-\slashed{\T},\slashed{\Phi}_+ \rbrack-4 e^{-A}\re (\mu\ \ov{\slashed{\Phi}_-}) \ . \nn
\eea
We rewrite the $R$-flux terms via the above rules: denoting $\te^a \w \pm \eta^{ab} \iota_b$ by $\te \pm \iota$, we get
\bea
&-\frac{1}{4}\lbrack \slashed{R},\slashed{\Phi}_+\rbrack -\frac{1}{8} \eta_{ad}\g^a \tilde{\slashed{\Phi}}_+ \g^{bc} \eta_{be} \eta_{cf} R^{def} +\frac{1}{8}\eta_{ad} \eta_{be} \eta_{cf} R^{def} \g^{bc} \slashed{\Phi}_+  \g^a \label{anticomR}\\
=&-\frac{1}{8}\eta_{ad} \eta_{be} \eta_{cf} R^{def}\left(\frac{1}{3}\left( (\te+\iota)^3 - (\te-\iota)^3  \right) + (\te+\iota)(\te-\iota)^2 -(\te+\iota)^2(\te-\iota) \right)^{abc} \Phi_+\nn\\
=&-\frac{1}{3}R^{abc}\iota_{a}\iota_{b}\iota_{c} \Phi_+ \nn \ ,
\eea
where the last two lines should be overall slashed. $\te$ and $\iota$ with different indices anticommute, and the $R$-flux is antisymmetric, so maintaining the indices $a,b,c$ fixed allows to commute $\te$ and $\iota$ in the second line. With notations of the Introduction and appendix \ref{ap:conv}, we finally obtain
\bea
2\!\!\!\!\! \begin{picture}(10,0)(-10,2)\put(0,0){\line(2,1){20}}\end{picture}{ \d \P_+}=& \lbrack \g^a,\cN_a \slashed{\Phi}_+\rbrack -\lbrace \slashed{\del}\left(2A-\tp\right), \slashed{\Phi}_+\rbrace + \lbrack \slashed{\b}_{\del}\left(2A+\tp\right)-\slashed{\T},\slashed{\Phi}_+ \rbrack-2\!\!\!\!\! \begin{picture}(10,0)(-10,2)\put(0,0){\line(4,1){42}}\end{picture}{ \ R \vee\P_+}-4 e^{-A}\ \i \hspace{-.2cm}\begin{picture}(10,10)(-10,5)
\put(0,2){\line(4,1){45}} \end{picture} \im(\overline{\mu} \Phi_-) \nn \\
= &2 \ \left( \cN^a \cdot \iota_a - \del_a\left(2A-\tp\right) \te^a \w + \left(\b^{ab}\del_b\left(2A+\tp\right)-\T^a\right) \iota_a - R \vee\right)\P_+ -4 e^{-A}\ \i \im(\overline{\mu} \Phi_-) \ ,\nn
\eea
where the last line should be overall slashed. We compute similarly
\bea
2\!\!\!\!\! \begin{picture}(10,0)(-10,2)\put(0,0){\line(2,1){20}}\end{picture}{ \d\P_-} =& \lbrack \g^a, \N_a \slashed{\Phi}_-\rbrack \label{dP-Bcom}\\
=& \slashed{\N} \eta^{1}_+ \otimes \eta^{2 \dag}_- + \g^a \eta^{1}_+ \otimes \left(\N_a\eta^{2}_-\right)^{\dagger}  - \N_a \eta^{1}_+ \otimes \eta^{2 \dag}_- \g^a -\eta^{1}_+ \otimes \left(\slashed{\N}\eta^{2}_-\right)^{\dagger} \nn\\
=& \left(\left(\slashed{\cN}  - \frac{1}{4} \slashed{R} -\slashed{\del}\left(2A-\tp\right)+ \slashed{\b}_{\del}\left(2A+\tp\right)-\slashed{\T}\right) \eta^{1}_+ -2e^{-A}\mu\eta^{1}_-\right)\otimes \eta^{2 \dag}_-\nn\\
+& \g^a \eta^{1}_+ \otimes \left(\eta_{ab}\cN^b \eta^{2}_- +\frac{1}{8} \eta_{ad} \eta_{be} \eta_{cf} R^{def} \g^{bc}\eta^{2}_-\right)^{\dagger}\nn \\
-& \left(- \eta_{ab}\cN^b \eta^{1}_+ +\frac{1}{8}\eta_{ad} \eta_{be} \eta_{cf} R^{def} \g^{bc} \eta^{1}_+\right) \otimes \eta^{2 \dag}_- \g^a\nn\\
-&\eta^{1}_+ \otimes \left(\left(- \slashed{\cN}  - \frac{1}{4} \slashed{R}-\slashed{\del}\left(2A-\tp\right)- \slashed{\b}_{\del}\left(2A +\tp\right) +\slashed{\T}\right) \eta^{2}_- +2e^{-A}\ov{\mu} \eta^{2}_+\right)^{\dagger}\nn\\
=&\lbrace \g^a,\cN_a \slashed{\Phi}_-\rbrace -\frac{1}{8}\left(2 \lbrace \slashed{R},\slashed{\Phi}_-\rbrace + \eta_{ad}\g^a \slashed{\Phi}_- \g^{bc} \eta_{be} \eta_{cf} R^{def} + \eta_{ad}  \eta_{be} \eta_{cf} R^{def} \g^{bc} \slashed{\Phi}_- \g^a\right)\nn\\
-&\lbrack \slashed{\del}\left(2A-\tp\right), \slashed{\Phi}_-\rbrack + \lbrace \slashed{\b}_{\del}\left(2A + \tp\right)-\slashed{\T}, \slashed{\Phi}_- \rbrace-4e^{-A}\mu \re \left(\slashed{\Phi}_+\right)\nn \ ,
\eea
and as above
\bea
& -\frac{1}{8}\left(2 \lbrace \slashed{R},\slashed{\Phi}_-\rbrace + \eta_{ad}\g^a \slashed{\Phi}_- \g^{bc} \eta_{be} \eta_{cf} R^{def} + \eta_{ad}  \eta_{be} \eta_{cf} R^{def} \g^{bc} \slashed{\Phi}_- \g^a\right) \label{comR}\\
=&-\frac{1}{8} \eta_{ad}  \eta_{be} \eta_{cf} R^{def} \left(\frac{1}{3}\left( (\te+\iota)^3 - (\te-\iota)^3  \right) + (\te+\iota)(\te-\iota)^2 -(\te+\iota)^2(\te-\iota) \right)^{abc} \Phi_-\nn\\
=&-\frac{1}{3}R^{abc}\iota_{a}\iota_{b}\iota_{c}\Phi_- \nn \ ,
\eea
where the last two lines should be slashed, as well as the following resulting one
\bea
2 \d \P_- =& 2\left( \cN^a \cdot \iota_a - \del_a\left(2A-\tp\right) \te^a \w + \left(\b^{ab}\del_b\left(2A+\tp\right)-\T^a\right) \iota_a  - R \vee\right)\P_- -4e^{-A}\mu \re \left(\Phi_+\right) \ .\nn
\eea
Using the Clifford map backwards, we finally get for type IIB two equations on forms
\bea
\!\!\!\!\!\!\!\!\!\!\!\!\!\!\! e^{\tp} \left(\d -  \cN^a \cdot \iota_a+\T \vee + R \vee\right) \left(e^{-\tp} \P_+\right) + e^{-2A}\left( \d  + \cN^a \cdot \iota_a \right)(e^{2A}) \P_+ &=-2e^{-A}\ \i \im(\overline{\mu} \Phi_-) \label{dP+0B} \\
\!\!\!\!\!\!\!\!\!\!\!\!\!\!\! e^{\tp} \left(\d -  \cN^a \cdot \iota_a+\T \vee + R \vee\right) \left(e^{-\tp} \P_-\right) + e^{-2A}\left( \d  + \cN^a \cdot \iota_a \right)(e^{2A}) \P_- &=-2e^{-A} \mu \re(\Phi_+) \ . \label{dP-0B}
\eea
This calculation can be done as well in type IIA: the difference comes from the SUSY conditions \eqref{susy2} and \eqref{susy3} where one has to change the chirality of $\eta^2$. The above computation can be reproduced almost identically considering $\d \P_+$ in place of $\d \P_-$ and vice versa: this replaces $\eta^2_{\pm}$ into one another, and the type IIA SUSY conditions can then be used, leading simply to an exchange of $\P_+$ and $\P_-$ in the computation. Doing so, commutators and anti-commutators get exchanged because of the even/odd degree change, but this goes through without issue; in particular we get eventually the same $R$-flux term, since \eqref{anticomR} and \eqref{comR} give the same resulting action on the pure spinors. The only difference in the process may appear in the $\mu$-terms, because of \eqref{signslash}, and in the signs induced by the (anti)-commutators. In the end, we obtain in type IIA
\bea
\!\!\!\!\!\!\!\!\!\!\!\! e^{\tp}\left(\d -  \cN^a \cdot \iota_a+\T \vee + R \vee\right) \left(e^{-\tp} \P_-\right) + e^{-2A}\left( \d  + \cN^a \cdot \iota_a \right)(e^{2A}) \P_- & =2e^{-A}\ \i \im(\overline{\mu} \Phi_+) \label{dP-0A} \\
\!\!\!\!\!\!\!\!\!\!\!\! e^{\tp}\left(\d -  \cN^a \cdot \iota_a+\T \vee + R \vee\right) \left(e^{-\tp} \P_+\right) + e^{-2A}\left( \d  + \cN^a \cdot \iota_a \right)(e^{2A}) \P_+ & =2e^{-A} \mu\ \re(\Phi_-) \ . \label{dP+0A}
\eea

The above computation should in principle be completed by RR contributions, that we do not have here; we would still like to obtain the result as if they were present. We thus follow closely the analogous computation done for standard supergravity with RR fluxes in \cite{Grana:2006kf} (the result is even specified there not to hold without RR), and we perform an additional step, which in absence of RR may not look required. It involves the SUSY condition \eqref{susy1}, that has not been used so far. From that condition in type IIB, we obtain
\bea
& 0=\eta_-^1 \otimes \left(\mu\ \eta_-^{2} + e^{A} \left( \s{\del} A -  \s{\b}_{\del} A \right) \eta_+^{2} \right)^{\dag *}= \mu \ov{\s{\P}_-}- e^{A} \ \ov{\s{\P}_+}  \left( \s{\del} A -  \s{\b}_{\del} A \right) \\
& 0=\left(\mu\ \eta_-^{1} + e^{A} \left( \s{\del} A + \s{\b}_{\del} A \right) \eta_+^{1} \right)^{*} \otimes \eta_-^{2 \dag} = \ov{\mu} \s{\P}_- - e^{A} \left( \s{\del} A + \s{\b}_{\del} A \right) \ov{\s{\P}_+} \ ,
\eea
from which we deduce
\bea
& 0= 2\re ( \ov{\mu} \s{\P}_-) - e^A \lbrace \s{\del} A , \ov{\s{\P}_+} \rbrace - e^A \lbrack \s{\b}_{\del} A , \ov{\s{\P}_+} \rbrack \\
\longleftrightarrow \quad & 0=e^{-A}\ \i \im(\overline{\mu} \Phi_-) - e^{-A}\left( \d  - \cN^a \cdot \iota_a \right)(e^{A}) \ov{\P_+ } \ . \label{addedpiece}
\eea
We subtract this quantity on the RHS of \eqref{dP+0B} and get
\bea
e^{\tp}\left(\d -  \cN^a \cdot \iota_a+\T \vee + R \vee\right) \left(e^{-\tp} \P_+\right) +& e^{-2A}\left( \d  + \cN^a \cdot \iota_a \right)(e^{2A}) \P_+ \label{dP+1} \\
&  =-3e^{-A}\ \i \im(\overline{\mu} \Phi_-) + e^{-A}\left( \d  - \cN^a \cdot \iota_a \right)(e^{A}) \ov{\P_+ } \ . \nn
\eea
In type IIA, we proceed similarly with the SUSY condition \eqref{susy1}
\bea
& 0=\eta_-^1 \otimes \left(\mu\ \eta_+^{2} + e^{A} \left( \s{\del} A -  \s{\b}_{\del} A \right) \eta_-^{2} \right)^{\dag *}= \mu \ov{\s{\P}_+}- e^{A} \ \ov{\s{\P}_-}  \left( \s{\del} A -  \s{\b}_{\del} A \right) \\
& 0=\left(\mu\ \eta_-^{1} + e^{A} \left( \s{\del} A + \s{\b}_{\del} A \right) \eta_+^{1} \right)^{*} \otimes \eta_+^{2 \dag} = \ov{\mu} \s{\P}_+ - e^{A} \left( \s{\del} A + \s{\b}_{\del} A \right) \ov{\s{\P}_-} \ ,
\eea
to get
\bea
& 0= 2\ \i \im ( \ov{\mu} \s{\P}_+) - e^A \lbrack \s{\del} A , \ov{\s{\P}_-}  \rbrack - e^A \lbrace \s{\b}_{\del} A , \ov{\s{\P}_-} \rbrace \\
\longleftrightarrow \quad & 0=e^{-A}\ \i \im(\overline{\mu} \Phi_+) + e^{-A}\left( \d  - \cN^a \cdot \iota_a \right)(e^{A}) \ov{\P_- } \ .
\eea
We add this quantity to the RHS of \eqref{dP-0A} to obtain
\bea
e^{\tp}\left(\d -  \cN^a \cdot \iota_a+\T \vee + R \vee\right) \left(e^{-\tp} \P_-\right) +& e^{-2A}\left( \d  + \cN^a \cdot \iota_a \right)(e^{2A}) \P_- \label{dP-1} \\
&  =3e^{-A}\ \i \im(\overline{\mu} \Phi_+) + e^{-A}\left( \d  - \cN^a \cdot \iota_a \right)(e^{A}) \ov{\P_- } \ . \nn
\eea
There is a priori no RR contribution to the other pure spinor condition, so we do not modify \eqref{dP-0B} or \eqref{dP+0A}. Our final pure spinors conditions are then given by \eqref{dP-0B} and \eqref{dP+1} in type IIB, and \eqref{dP+0A} and \eqref{dP-1} in type IIA, as summarized in \eqref{dP1} and \eqref{dP2}.

\subsection{On the sufficiency of the pure spinors conditions}\label{ap:suff}

In section \ref{sec:purespin} and appendix \ref{ap:purespin}, we have derived the pure spinors conditions \eqref{dP1} and \eqref{dP2} using the SUSY conditions \eqref{susy1}, \eqref{susy2} and \eqref{susy3}; in other words, we have shown that \eqref{dP1} and \eqref{dP2} are necessary for the backgrounds of interest to preserve SUSY. We study here whether these two conditions are also sufficient. Following \cite{Grana:2006kf}, this amounts to considering a generic expansion of $\N_a \eta_+^i$ and of further quantities appearing in \eqref{susy1}, \eqref{susy2} and \eqref{susy3} on a complete basis of six-dimensional spinors. One then checks whether the coefficients in these expansions are determined by the pure spinors conditions to be those of the SUSY conditions. It will turn out not to be the case, implying that the conditions \eqref{dP1} and \eqref{dP2} are not sufficient. We argue that this is due to the absence of RR fluxes.

We start by expanding the following combinations on a complete basis of spinors $\{\eta^{1,2}$, $\g^{a}\eta^{1,2}$, $\g_{(6)}\eta^{1,2}\}$. Taking chiralities into account, we get in type IIB
\bea
\!\!\!\!\!\!\!\! \Big(\g^a\left(\N_a  \mp \eta_{ad} \cN^d\right)+\frac{1}{24} \eta_{ad} \eta_{be} \eta_{cf} R^{def} \g^{abc}\Big)\eta_+^{1,2}&=\Big(T^{1,2}+\i U^{1,2}\g_{(6)} \Big)\ \eta_-^{1,2} + V_a^{1,2}\g^a \eta_+^{1,2} \label{genDexp}\\
\!\!\!\!\!\!\!\! \left( \N_{a} \pm \eta_{ad} \cN^{d} - \frac{1}{8} \eta_{ad} \eta_{be} \eta_{cf} R^{def} \g^{bc} \right) \eta^{1,2}_+ &=\Big(P_a^{1,2}+\i Q_a^{1,2}\g_{(6)} \Big)\ \eta_+^{1,2} + \i S_{ad}^{1,2}\g^d \eta_-^{1,2}  \ ,\nn
\eea
where the coefficients $V_a^{1,2}$, $P_a^{1,2}$ and $Q_a^{1,2}$ must be real. A more generic situation would be to consider only $\nabla$ on the internal spinors. However \eqref{dP1} and \eqref{dP2} impose without ambiguity these particular combinations of $\N$, $\cN$ and $R$-flux to act on the spinors, so there is actually no restriction here. From these generic expansions, we compute the exterior derivative of the pure spinors as in \eqref{dP+Bcom} and \eqref{dP-Bcom}
\bea
2\!\!\!\!\! \begin{picture}(10,0)(-10,2)\put(0,0){\line(2,1){20}}\end{picture}{\d\P_+} =& \{ \g^a, \N_a \slashed{\P}_+\} \label{dP+gen}\\
=& \slashed{\N} \eta^{1}_+ \otimes \eta^{2 \dag}_+ + \g^a \eta^{1}_+ \otimes \left(\N_a\eta^{2}_+\right)^{\dag}  + \N_a \eta^{1}_+ \otimes \eta^{2 \dag}_+ \g^a +\eta^{1}_+ \otimes \left(\slashed{\N}\eta^{2}_+\right)^{\dag} \nn\\
=& \left(\left(\slashed{\cN}  - \frac{1}{4} \slashed{R}+\slashed{V}^{1} \right) \eta^{1}_+ +\Big(T^{1}+\i U^{1}\g_{(6)} \Big)\eta^{1}_-\right)\otimes \eta^{2 \dag}_+\nn\\
+& \g^a \eta^{1}_+ \otimes \left(\eta_{ab}\cN^b \eta^{2}_+ +\frac{1}{8}\eta_{ad} \eta_{be} \eta_{cf} R^{def} \g^{bc}\eta^{2}_+ + \Big(P_a^{2}+\i Q_a^{2}\g_{(6)} \Big)\ \eta_+^{2} + \i S_{ad}^{2}\g^d \eta_-^{2}  \right)^{\dag} \nn\\
+& \left(- \eta_{ab}\cN^b \eta^{1}_+ +\frac{1}{8}\eta_{ad} \eta_{be} \eta_{cf} R^{def} \g^{bc}\eta^{1}_+ + \Big(P_a^{1}+\i Q_a^{1}\g_{(6)} \Big)\ \eta_+^{1} + \i S_{ad}^{1}\g^d \eta_-^{1} \right) \otimes \eta^{2 \dag}_+ \g^a\nn\\
+&\eta^{1}_+ \otimes \left(\left(- \slashed{\cN}  - \frac{1}{4} \slashed{R}+\slashed{V}^{2} \right) \eta^{2}_+ +\Big(T^{2}+\i U^{2}\g_{(6)} \Big)\ \eta_-^{2} \right)^{\dag}\nn\\
=& \lbrack \g^a,\cN_a \slashed{\Phi}_+\rbrack -2\!\!\!\!\! \begin{picture}(10,0)(-10,2)\put(0,0){\line(4,1){42}}\end{picture}{ \ R \vee\P_+}\nn\\
+& \g^a\slashed{\Phi}_+ \Big(P_a^{2}-\i Q_a^{2} \Big) - \i \ov{S_{ad}^{2}}\g^a \slashed{\Phi}_-\g^d + \Big(P_a^{1}+\i Q_a^{1} \Big)\ \sla{\P_+} \g^a + \i S_{ad}^{1}\g^d \ov{\slashed{\Phi}_-} \g^a\nn\\
+&\slashed{V}^{1}\slashed{\Phi}_+ +\slashed{\Phi}_+\slashed{V}^{2}+\Big(T^{1}-\i U^{1} \Big)\ \overline{\sla{\P_-}} + \sla{\P_-}\Big(\overline{T^{2}} + \i \overline{U^{2}} \Big)\nn\ ,
\eea
\bea
2\!\!\!\!\! \begin{picture}(10,0)(-10,2)\put(0,0){\line(2,1){20}}\end{picture}{ \d\P_-} =& \lbrack \g^a, \N_a \slashed{\Phi}_-\rbrack \label{dP-gen}\\
=& \slashed{\N} \eta^{1}_+ \otimes \eta^{2 \dag}_- + \g^a \eta^{1}_+ \otimes \left(\N_a\eta^{2}_-\right)^{\dagger}  - \N_a \eta^{1}_+ \otimes \eta^{2 \dag}_- \g^a -\eta^{1}_+ \otimes \left(\slashed{\N}\eta^{2}_-\right)^{\dagger} \nn\\
=& \left(\left(\slashed{\cN}  - \frac{1}{4} \slashed{R} +\slashed{V}^{1} \right) \eta^{1}_+ +\Big(T^{1}+\i U^{1}\g_{(6)} \Big)\eta^{1}_-\right)\otimes \eta^{2 \dag}_-\nn\\
+& \g^a \eta^{1}_+ \otimes \left(\eta_{ab}\cN^b \eta^{2}_- +\frac{1}{8} \eta_{ad} \eta_{be} \eta_{cf} R^{def} \g^{bc}\eta^{2}_- + \Big(P_a^{2}+\i Q_a^{2}\g_{(6)} \Big)\ \eta_-^{2} + \i \ov{S_{ad}^{2}}\g^d \eta_+^{2}\right)^{\dagger}\nn \\
-& \left(- \eta_{ab}\cN^b \eta^{1}_+ +\frac{1}{8}\eta_{ad} \eta_{be} \eta_{cf} R^{def} \g^{bc} \eta^{1}_+ + \Big(P_a^{1}+\i Q_a^{1}\g_{(6)} \Big)\ \eta_+^{1} + \i S_{ad}^{1}\g^d \eta_-^{1}\right) \otimes \eta^{2 \dag}_- \g^a\nn\\
-&\eta^{1}_+ \otimes \left(\left(- \slashed{\cN}  - \frac{1}{4} \slashed{R}+\slashed{V}^{2} \right) \eta^{2}_- -\Big(\overline{T^{2}}+\i \overline{U^{2}}\g_{(6)} \Big)\ \eta_+^{2}  \right)^{\dagger}\nn
\eea
\bea
2\!\!\!\!\! \begin{picture}(10,0)(-10,2)\put(0,0){\line(2,1){20}}\end{picture}{ \d\P_-}=&\lbrace \g^a,\cN_a \slashed{\Phi}_-\rbrace -2\!\!\!\!\! \begin{picture}(10,0)(-10,2)\put(0,0){\line(4,1){42}}\end{picture}{ \ R \vee\P_-}\nn\\
+& \g^a\slashed{\Phi}_- \Big(P_a^{2}+\i Q_a^{2} \Big) - \i S_{ad}^{2}\g^a \slashed{\Phi}_+\g^d - \Big(P_a^{1}+\i Q_a^{1} \Big)\ \sla{\P_-} \g^a - \i S_{ad}^{1}\g^d \ov{\slashed{\Phi}_+} \g^a\nn\\
+&\slashed{V}^{1}\slashed{\Phi}_- -\slashed{\Phi}_-\slashed{V}^{2}+\Big(T^{1}-\i U^{1} \Big)\ \overline{\sla{\P_+}} + \sla{\P_+}\Big(T^{2}-\i U^{2} \Big)\nn\ .
\eea
We then use the Clifford map on these equations. We first compare the result from \eqref{dP-gen} to \eqref{dP1} and deduce
\bea
& S_{ad}^{1}=S_{ad}^{2}=0, \ Q_a^{1}=Q_a^{2}=0\label{identif}\\
& P_a^2 + V_a^{1}=-\del_a\left(2A-\tp\right)+\eta_{ab}\left(\b^{bd}\del_d\left(2A + \tp\right)-\T^{b}\right)\nn\\
& P_a^1 + V_a^{2}=-\del_a\left(2A-\tp\right)-\eta_{ab}\left(\b^{bd}\del_d\left(2A + \tp\right)-\T^{b}\right)\nn\\
& T^{1}-\i U^1 = T^{2}-\i U^2 =-2e^{-A}\mu   \nn\ .
\eea
Fixing this way the coefficients reproduces \eqref{susy2} and \eqref{susy3}, provided one sets $P_a^1=P_a^2=0$; we will come back to that point. We turn to \eqref{dP+gen}: comparing it to \eqref{dP2}, taking into account the identifications \eqref{identif}, one obtains precisely \eqref{addedpiece} as a constraint. The latter should allow to reproduce the remaining SUSY condition \eqref{susy1}. To verify this, we introduce a generic expansion of the following quantity
\beq
\del_a A \g^a \eta_+^{1,2}= \g^a (R_a^{1,2} + \i W_a^{1,2}\g_{(6)}) \eta_+^{1,2}  - X^{1,2} \eta_-^{1,2} \ ,
\eeq
where $R_a^{1,2}$ and $W_a^{1,2}$ are real. Then, we consider the sum
\bea
0=&\eta_-^1 \otimes (X^{2} \eta_-^{2} + \del_a A \g^a \eta_+^{2}- \g^a (R_a^{2} + \i W_a^{2}\g_{(6)}) \eta_+^{2} )^{\dag *} \\
&+ (X^{1} \eta_-^{1} + \del_a A \g^a \eta_+^{1}- \g^a (R_a^{1} + \i W_a^{1}\g_{(6)}) \eta_+^{1} )^{*}  \otimes \eta_-^{2\dag} \nn\\
=& - \lbrace \s{\del} A , \ov{\s{\P}_+} \rbrace  + ( R_a^{2} + \i W_a^{2}) \ov{\s{\P}_+} \g^a  + ( R_a^{1} - \i W_a^{1}) \g^a \ov{\s{\P}_+} + \ov{\s{\P}_-} X^{2} + \s{\P}_- \ov{X^{1}} \ . \nn
\eea
Using the Clifford map on the last equation, and comparing the result to the obtained constraint \eqref{addedpiece}, we get
\beq
W^1_a=W^2_a=0\ ,\ R^1_a=-R^2_a=-\b^{ab}\del_b A \ ,\ X^1=X^2=\mu e^{-A} \ .
\eeq
This reproduces precisely the SUSY condition \eqref{susy1}.

To conclude, the SUSY conditions \eqref{susy1}, \eqref{susy2} and \eqref{susy3} are reproduced starting from the pure spinors conditions \eqref{dP1} and \eqref{dP2} in type IIB, provided one fixes $P_a^1=P_a^2=0$. The ambiguity or freedom in the $P_a^{1,2}$ is in our opinion related to the absence of RR fluxes: those would otherwise bring more constraints. The $P_a^{1,2}$ could also be related to the norms of the internal spinors, so far not needed. These norms are fixed in \cite{Grana:2006kf} thanks to the RR contributions; this may explain the ambiguity we get here. We conclude that the pure spinors conditions \eqref{dP1} and \eqref{dP2} are not sufficient, but the remaining ambiguity should be fixed by considering the RR sector. We expect the same situation in type IIA.

\section{The Dirac operator and the $\b$-twist}

\subsection{Rewriting of the Dirac operator}\label{ap:betatwist}

In this appendix we prove that the Dirac operator of $\b$-supergravity \eqref{Diraccurved} or \eqref{Diracflat} can be rewritten as \eqref{Debetaintro}, also given in \eqref{Debeta}. As explained in the Introduction and section \ref{sec:more}, the analogy with the twist by the $b$-field can lead to guessing this rewriting, as being the expression for a generalized derivative with non-geometric fluxes. For instance, $\b$-transforms acting in a spinorial representation as an exponential on pure spinors, as well as shifts on a background $Q$-flux, are mentioned in \cite{Micu:2007rd}. However, this common guess has never been proven: a first difficulty is to compute this operator \eqref{Debeta} where wedges and contractions mix, another one is the interpretation of such an operator at ten dimensions which was lacking before the introduction of $\b$-supergravity. Let us prove here this rewriting of the Dirac operator. To do so, we will first compute \eqref{Debeta} in curved indices and then rewrite it in flat indices to match it with \eqref{Diracflat}.

Contractions are anticommuting (see appendix \ref{ap:conv} for conventions), but $\b\vee$ contains an even number of them and is therefore commuting with itself. Powers of it are then naturally defined, and the exponential $e^{\b\vee}$ is defined by the serie. Standard properties are thus satisfied, namely $e^{\b\vee} e^{-\b\vee}=1$ and $\del_m e^{\b\vee}= e^{\b\vee} \del_m \b\vee$. We deduce for a $p$-form $A$
\beq
e^{\b\vee} \d (e^{-\b\vee} A)= e^{\b\vee} \d x^m \w e^{-\b\vee} \left(-\del_m (\b\vee) A + \del_m A \right) \ .
\eeq
Using the commutation relations of a one-form and a contraction, one can further verify
\bea
& \forall n \geq 1,\ \d x^m \w (\b\vee)^n= n (\b\vee)^{n-1} \b^{mr} \iota_r + (\b\vee)^n \d x^m \w\\
& \Rightarrow \d x^m \w e^{-\b\vee} = e^{-\b\vee} \d x^m \w - e^{-\b\vee} \b^{mr} \iota_r \ ,
\eea
from which we finally deduce
\beq
e^{\b\vee} \d (e^{-\b\vee} A)= \d A - \left(\d x^m \w \del_m (\b\vee) A + \b^{mr} \iota_r \del_m A \right) + \b^{mr} \iota_r \del_m (\b\vee) A \ .\label{Dcurved}
\eeq
The last term is proportional to the $R$-flux, thanks to the anticommuting properties of the contractions. We now rewrite the above expression with flat indices. Using
\beq
Q_c{}^{ab}= \te^q{}_c \te^a{}_{m} \te^b{}_{n} \left( \del_q \b^{mn} +2 \te^d{}_{q} \b^{r[m} \del_r \te^{n]}{}_{d} \right) \label{Qfluxeee} \ ,
\eeq
we obtain
\bea
e^{\b\vee} \d (e^{-\b\vee} A) &= \d A - \b^{ab} \iota_b \del_a \cdot A -\frac{1}{2} Q_a{}^{bc} \te^a\w \iota_b \iota_c  A + R\vee A\\
& - \frac{1}{(p-1)!} \b^{ab} \iota_b \te^m{}_c  \del_a (\te^{a_1}{}_m) A_{a_1 \dots a_p} \te^c \w \te^{a_2} \dots \te^{a_p} + \b^{db} \te^{c}{}_m \del_d (\te^m{}_a) \te^a\w \iota_b \iota_c  A \ ,\nn
\eea
where the dot stands for acting only on the form coefficient in flat indices. Thanks to the definition of $f^a{}_{bc}$, of a contraction, and the commutation relations, one can show
\bea
& - \frac{1}{(p-1)!} \b^{ab} \iota_b \te^m{}_c  \del_a (\te^{a_1}{}_m) A_{a_1 \dots a_p} \te^c \w \te^{a_2} \dots \te^{a_p}\\
&= \frac{1}{2} \b^{ab} f^c{}_{ab} \iota_c A + \b^{ab} \te^m{}_c  \del_a (\te^{a_1}{}_m) \te^c \w \iota_b \iota_{a_1} A \ ,\nn
\eea
from which we deduce
\beq
e^{\b\vee} \d (e^{-\b\vee} A) = \left(\d  - \b^{ab} \iota_b \del_a \cdot  -\frac{1}{2} Q_a{}^{bc} \te^a\w \iota_b \iota_c   + R\vee + \frac{1}{2} \b^{ab} f^c{}_{ab} \iota_c \right) A \ .
\eeq
Rescaling $A$, one gets further
\bea
& e^{\tp} e^{\b\vee} \d (e^{-\b\vee} e^{-\tp} A) \\
& = \left(\d  - \b^{ab} \iota_b \del_a \cdot  -\frac{1}{2} Q_a{}^{bc} \te^a\w \iota_b \iota_c   + R\vee + \frac{1}{2} \b^{ab} f^c{}_{ab} \iota_c - \del_a \tp \te^a\w + \b^{ab} \del_a \tp \iota_b  \right) A \ .\nn
\eea
This can be verified to be the same as one half of the Dirac operator \eqref{Diracflat}, recalling \eqref{TQ}. This proves \eqref{Debeta}.

\subsection{Invariance of the Dirac operator}\label{ap:invD}

We described in section \ref{sec:geomcharac} a class of geometric backgrounds of $\b$-supergravity, determined in \cite{Andriot:2014uda}, with well-defined global aspects. We further argued that it should be possible for them to construct $E_{T^*}$. Obtaining similar results beyond this subcase looks however unlikely, as we now argue. Inspired by standard supergravity and the role played by $b$-field gauge transformations for $E_T$, it seems a natural requirement to ask for the invariance of the Dirac operator \eqref{Debeta} when patching the fields. Let us determine accordingly the allowed transformations of $\b$. A transformation that admits an infinitesimal form can be written as the shift (assumed antisymmetric)
\beq
\delta \b^{pq} = \varpi^{pq} \ .\label{transfo}
\eeq
In addition, we consider that it only acts on $\b$, and not for instance on the derivatives, so we do not include diffeomorphisms here. We use for the Dirac operator $\D$ the simple expression in curved indices \eqref{Dcurved} (we neglect the dilaton that does not play a role here). For $\D$ to stay invariant under \eqref{transfo}, one should have for any form $A$
\beq
- \left(\d x^m \w \del_m (\varpi\vee) A + \varpi^{mr} \iota_r \del_m A \right) + \delta\left(\b^{mr} \iota_r \del_m (\b\vee)\right) A =0 \ .
\eeq
This polyform equation should vanish degree by degree, so the last term vanishes independently of the first two. Since this should hold for any form $A$, we can first consider constant ones, implying
\bea
& \d x^m \w \del_m (\varpi\vee) A =0 \quad \forall A \ \mbox{constant}\\
\Leftrightarrow\ & \forall m,\ \del_m (\varpi^{pq}) \iota_p \iota_q A =0 \quad \forall A \ \mbox{constant}\\
\Leftrightarrow\ & \forall m,p,q,\ \del_m (\varpi^{pq}) =0 \ .\label{cond1}
\eea
We deduce that for any form $A$
\beq
\varpi^{mr} \iota_r \del_m A = 0 \Leftrightarrow  \forall r,\ \varpi^{rm} \del_m X = 0 \ ,\label{cond2}
\eeq
where $X$ stands for any tensor component. For example, the components of the metric can be used into the components of some form $A$, so $\varpi^{rm} \del_m$ should vanish on them. Conditions \eqref{cond1} and \eqref{cond2} imply
\beq
\delta\left(\b^{mr} \iota_r \del_m (\b\vee)\right) = 0 \ ,
\eeq
so these two conditions are also sufficient to verify the invariance of $\D$.

The requirement can be refined by asking the invariance of $\D$ not on any form $A$, but for forms whose components are built from physical tensors and fields (e.g. the pure spinors). In addition, asking for the presence of $n$ isometries allows to solve the condition \eqref{cond2} by having $\varpi^{pq}$ possibly non-zero only along the isometries (see also \cite{Andriot:2014uda} for a relation between \eqref{cond2} and isometries). The transformation \eqref{transfo} then boils down to the $\b$-transform \eqref{betatransfo}, bringing us back to the subcase mentioned above. Going beyond seems however difficult: for instance, allowing for a priori any linear coordinate dependence in $X$ implies $\varpi=0$.

\end{appendix}

\newpage

\providecommand{\href}[2]{#2}\begingroup\raggedright

\endgroup

\end{document}